# Reconciling Magma-Ocean Crystallization Models with the Present-day Structure of the Earth's mantle

Maxim D. Ballmer[1,2]*, Diogo L. Lourenço[1], Kei Hirose[2], Razvan Caracas[3], and Ryuichi Nomura[2,4]

(1) Institut für Geophysik, ETH Zürich, 8092 Zurich, Switzerland
(2) Earth-Life Science Institute, Tokyo Institute of Technology, Meguro, Tokyo 152-8550, Japan
(3) Laboratoire de Géologie de Lyon UMR CNRS 5276, Ecole Normale Supérieure de Lyon, 69364 Lyon, France
(4) Geodynamics Research Center, Ehime University, Matsuyama 790-8577, Japan
(*) corr. author: maxim.ballmer@erdw.ethz.ch

## Abstract

Terrestrial planets are thought to experience episode(s) of large-scale melting early in their history. Fractionation during magma-ocean freezing leads to unstable stratification within the related cumulate layers due to progressive iron enrichment upwards, but the effects of incremental cumulate overturns during MO crystallization remain to be explored. Here, we use geodynamic models with a moving-boundary approach to study convection and mixing within the growing cumulate layer, and thereafter within the fully-crystallized mantle. For fractional crystallization, cumulates are efficiently stirred due to subsequent incremental overturns, except for strongly iron-enriched late-stage cumulates, which persist as a stably stratified layer at the base of the mantle for billions of years. Less extreme crystallization scenarios can lead to somewhat more subtle stratification. In any case, the long-term preservation of at least a thin layer of extremely enriched cumulates with Fe#>0.4, as predicted by all our models, is inconsistent with seismic constraints. Based on scaling relationships, however, we infer that final-stage Fe-rich magma-ocean cumulates originally formed near the surface should have overturned as small diapirs, and hence undergone melting and reaction with the host rock during sinking. The resulting moderately iron-enriched metasomatized/hybrid rock assemblages should have accumulated at the base of the mantle, potentially fed an intermittent basal magma ocean, and be preserved through the present-day. Such moderately iron-enriched rock assemblages can reconcile the physical properties of the large low shear-wave velocity provinces in the present-day lower mantle. Thus, we reveal Hadean melting and rock-reaction processes by integrating magma-ocean crystallization models with the seismic-tomography snapshot.

## 1. Introduction

Due to massive energy release during planetary accretion and differentiation, terrestrial planets typically go through an episode (or multiple episodes) of large-scale melting very early in their history. In these episodes, not only the metallic core, but also the rocky mantle is thought to be largely molten to form a deep and global magma ocean (MO). The main factors to sustain planetary-scale melting are gravitational energy release during core formation (*Flasar and Birch*, 1973; *Sasaki and Nakazawa*, 1986), radioactive heating by short-lived isotopes (*Mostefaoui et al.*, 2005; *Urey*, 1955), thermal blanketing due to the presence of an early atmosphere (*Abe and Matsui*, 1986; *Genda and Abe*, 2003; *Hamano et al.*, 2013; *Lebrun et al.*, 2013) and shock heating due to giant impacts (*Nakajima and Stevenson*, 2015). For the Earth, the proposed moon-forming giant impact and related core segregation has provided the energy for perhaps the last and most intensive MO episode (*Canup*, 2012; *Ćuk and Stewart*, 2012; *Nakajima and Stevenson*, 2015; *Tonks and Melosh*, 1993). This last major MO episode has set the initial condition for the evolution of a tectonically active planet that has been able to sustain surface water, a dense atmosphere, and abundant life for billions of years.

Despite these important implications for Earth evolution, key aspects of magma-ocean cooling and crystallization remain poorly understood. For example, it remains unclear whether the MO has frozen from the bottom upwards (*Abe*, 1997; *Solomatov*, 2000), or started freezing at intermediate depths (*Labrosse et al.*, 2007; *Nomura et al.*, 2011; *Stixrude et al.*, 2009). This first-order question has important implications for planetary thermochemical evolution, because freezing from mid-mantle depths outwards implies that a basal magma ocean (BMO) is chemically and thermally isolated in the deep mantle (*Labrosse et al.*, 2007). The depth range of possible BMO formation depends on the poorly-constrained crystallization behavior of silicate liquids at high pressures (i.e., the relationship between the solidus and adiabiat) (*Fiquet et al.*, 2010; *Mosenfelder et al.*, 2009; *Stixrude et al.*, 2009), as well as the density difference between crystals and liquids as a function of depth (*Andrault et al.*, 2017). In any case, the surficial MO will ultimately evolve independently of the (putatively present) BMO.

Another key unresolved question concerns the timescales of MO crystallization, which have been proposed to range from thousands (*Solomatov*, 2000), to millions of years (*Abe*, 1997; *Lebrun et al.*, 2013). Longer timescales of MO freezing generally imply relatively slow convection within the MO and may allow for crystal settling and fractional crystallization (*Solomatov*, 2015), depending on magma viscosity (*Solomatov and Stevenson*, 1993b; *Tonks and Melosh*, 1990). On the other hand, short timescales imply fast turbulent convection, and hence equilibrium crystallization. MO crystallization timescales strongly depend on the nature and thickness of the proto-atmosphere, which ultimately controls the heat flux from Earth to space (*Lebrun et al.*, 2013; *Marcq*, 2012). As long as some degree of chemical fractionation occurs and the MO freezes from the bottom upwards, the MO is predicted to become progressively Fe-rich as Mg-rich bridgmanite in the lower mantle (*Andrault et al.*, 2012; *Boukare et al.*, 2015; *Fiquet et al.*, 2010; *Nomura et al.*, 2011; *Tateno et al.*, 2014), and other Mg-rich phases (mostly olivine and pyroxenes) in the upper mantle, are successively removed. The related Fe-enrichment in the coexisting cumulates leads to unstable compositional density stratification in the early mantle (*Elkins-Tanton*, 2008). Convective overturn of any unstable stratification ultimately results in stable compositional stratification, with a degree of stability that would depend on the degree of chemical fractionation during freezing, and hence the timescale of freezing.

Thus, magma-ocean processes can lead to two fundamentally different paths of planetary evolution. First, stable compositional stratification of the mantle after cumulate overturn may be too strong to be erased by subsequent thermally-powered solid-state convection, and layered mantle convection with little or no material exchange across layers may persist for the age of the solar system (*Plesa et al.*, 2014; *Tosi et al.*, 2013). Second, stratification may be sufficiently moderate, at least across most of the mantle, to allow for whole-mantle convection and mixing. By imaging subducted slabs deep in the mantle, which attest to whole-mantle convection and stirring, seismic tomography indicates that the latter is true on planet Earth (*Grand et al.*, 1997; *van der Hilst et al.*, 1997; *van der Meer et al.*, 2010). Although compositional heterogeneity has been identified in the lowermost mantle and is potentially linked to MO freezing processes (*Garnero and McNamara*, 2008; *Hernlund and McNamara*, 2015; *Tackley*, 2012), it does not exceed ~3% of the mantle's volume (*Hernlund and Houser*, 2008).

In the absence of geological and geophysical data from the Archean eon, evidence for a largely well-mixed modern mantle can be used as a constraint for MO evolution. For the Earth, any MO crystallization scenarios leading to persistent mantle stratification that is inconsistent with the structure of the present-day mantle can be ruled out. Pronounced stratification is expected for a single overturn of the compositionally stratified cumulate layers (*Elkins-Tanton et al.*, 2005; *Elkins-Tanton*, 2008). In turn, initial stratification may be (partially) erased through progressive mixing due to multiple incremental cumulate overturns. The timescales of Rayleigh-Taylor instability in the cumulate layers are indeed comparable with those of MO freezing (*Solomatov and Stevenson*, 1993a). However, the consequences of cumulate mixing due to incremental overturns for Earth evolution have not yet been explicitly modeled.

In this study, we use geodynamic models to explore compositional fractionation and cumulate overturn during MO crystallization, as well as the related consequences for subsequent mantle thermochemical evolution. Convection is modeled in the growing cumulate layer (i.e. during MO freezing), and is continued to be modeled in the fully solid state for billions of years to address long-term mantle evolution. Ultimately, we compare the predictions of our models in terms of the final extent of mantle heterogeneity with present-day geophysical and geochemical observations.

## 2. Methods

To model cumulate overturn during MO crystallization and subsequent thermochemical evolution of the mantle, we solve the conservation equations of mass, momentum and energy in a two-dimensional Cartesian box of aspect ratio one (i.e., with both width and depth $x_{box} = z_{box} =$ 2900 km) using an extended version of the finite-element code CitcomCU (*Moresi et al.*, 1996; *Zhong et al.*, 2000). Regarding the uncertainties in mineral-physics data for melt and rock properties at high pressures, the lack of geologic constraints for the Hadean eon (e.g. in terms of atmospheric composition (*Zahnle et al.*, 2010)), and last but not least, the challenges of explicitly modeling the dynamics of global-scale two-phase (melt-solid) dynamics, we make various simplifications. For example, our approach implies that the whole mantle is initially molten (*Canup*, 2012; *Ćuk and Stewart*, 2012; *Nakajima and Stevenson*, 2015). Note however that our cases in scenarios (EI) and (EIF) can be also interpreted in such a way that the depth extent of the MO did not exceed 1450 km (see below). We also assume that the MO is compositionally homogeneous (pyrolitic), and freezes from the bottom up (i.e., the pyrolite solidus and liquidus are steeper than the adiabat at all mantle depths (*Thomas et al.*, 2012; *Zhang and Herzberg*, 1994)). We further consider that any non-adiabatic thermal variations within the vigorously convecting MO are negligible (*Abe*, 1997; *Solomatov*, 2000; *Tonks and Melosh*, 1990). Thus, we set the (potential) temperature of all computational nodes within the MO to $T_{MO}$, with $T_{MO}$ evolving over time in a pre-defined fashion. We solve for viscous flow only within the cumulate layer, and set velocities within the MO to zero. Accordingly, the magma-ocean domain within the model box acts solely as a boundary condition for the cumulate domain (see Fig. 1). The boundary between the two domains moves with model time as the MO cools, according to the freezing temperatures of the MO as a function of MO composition (see below). MO cooling is pre-defined in such a way that the boundary moves at a constant velocity $v_{MOC}$, setting the characteristic time-scale of magma-ocean crystallization as $\tau_{MOC} = z_{box} / v_{MOC}$. At $t > \tau_{MOC}$, i.e. once the MO is completely crystallized, the simulation continues as a "normal" mantle-convection simulation. Depending on the crystallization scenario, some cases with fast MO freezing, assume $\tau_{MOC} = 0$ (see below).

The conditions imposed at all side boundaries as well as at the internal MO-cumulate domain boundary involve a free-slip criterion: stresses parallel to any boundary cannot be supported, and neither in- nor outflow are allowed to occur. Temperatures are set to $T_{MO}$ at the internal domain boundary (and within the overlying MO) and to $T_{CMB}$ at the bottom. As soon as the MO is fully solidified (i.e., for model time $t \geq \tau_{MOC}$), the top boundary condition is set to T = 0 °C. Initial conditions involve a cumulate layer that extends from the CMB to $d_{MO,init}$; the MO makes up the rest of the domain. $d_{MO,init}$ is usually 2610 km (or 290 km above the CMB), but depends on the MO crystallization scenario (see below). For example, for end-member fast-freezing cases with $\tau_{MOC} = 0$, $d_{MO,init}$ is also implicitly zero.

The initial thermal and compositional profiles of the cumulate layer are computed from a one-dimensional semi-analytical model of MO crystallization and related MO compositional evolution (Fig. 2), assuming that no cumulate overturn occurs during precipitation of the initial cumulate layer (of thickness $z_{box} - d_{MO,init}$) at the base of the mantle (note that the effective Rayleigh number scales with the thickness of the cumulate layer cubed). The resolution of the mesh is ~11 km, but the vertical resolution is increased to 7.25 km in the top 580 km of the box. The compositional profile of the initial cumulate layer, as well as that of each subsequent (incremental) addition to the cumulate during MO compositional evolution, are calculated from melt-solid equilibria during end-member equilibrium, fractional and "intermediate" crystallization, as well as combined scenarios, in a simplified $MgSiO_3$-$FeSiO_3$ compositional system (for details, see section 3). We assume a constant partitioning coefficient $K_D = 0.3$ (e.g., *Corgne et al.*, 2005; *Liebske et al.*, 2005; *Takahashi and Kushiro,* 1983), where:

$$K_D = \frac{Fe_{solid} Mg_{liquid}}{Mg_{solid} Fe_{liquid}}$$

with $Mg_{solid}$ and $Mg_{liquid}$ as well as $Fe_{solid}$ and $Fe_{liquid}$ being molar Mg and Fe concentrations in the solid and liquid, respectively. The starting composition is similar to pyrolite in terms of the molar iron to magnesium ratio (of 0.12:1). Si and O concentrations are chosen such that they are ultimately consumed as crystallization of $(Mg,Fe)SiO_3$ is completed and the MO is fully solidified. Initial compositional profiles are shown in Figure 2.

The initial thermal profile of the cumulate layer is defined by the MO freezing temperatures as a function of MO iron content and depth. The temperature of each infinitesimally thin cumulate layer that is precipitated from the MO is bracketed by the solidus and the liquidus of the MO (fractional-crystallization cases), or by the solidus and the 50%-solidus (intermediate-crystallization and combined scenarios). We explore both these upper-bound and lower-bound thermal scenarios. Pyrolite solidus and liquidus are parameterized to fit the experimental data from *Fiquet et al.* (2010). We take the pyrolite solidus to linearly increase and decrease from 3300 °C at $z$ = 1740 km to 4000 °C at the CMB and 1740 °C at the surface, respectively. This bi-linear fit to experimental constraints is a crude simplification (*Andrault et al.*, 2011; *Du and Lee*, 2014), but our conclusions do not depend on the choice of the melting curve (see section 4). The pyrolite liquidus is parabolically fitted as $T_{liq}$ = 1761.1 °C + $A \cdot z$ + $B \cdot z^2$, with $A$ = 1.971 K/km and $B$ = -0.00029396 K/km$^2$. The solidus as well as the liquidus are linearly depressed as a function of MO iron content by $k(Fe\#_{MO} - Fe\#_{PYR})$ with $k$ = 784 K, $Fe\#_{PYR}$ = 0.12/1.12 ≈ 0.107 and $Fe\#_{MO}$ = $Fe_{liquid}/(Mg_{liquid} + Fe_{liquid})$. Accordingly, the minimum solidus temperature that can occur for maximal $Fe\#_{MO}$ at the surface is 1000 °C. As core cooling is significantly inhibited as soon as the first cumulate layer is precipitated, $T_{CMB}$ is taken to be 4000 °C in the lower-bound (solidus) thermal scenario and 4500 °C in the upper-bound thermal scenario. Note that all the values reported above (in °C) are "real" temperatures; the adiabatic gradient $\gamma$ is removed from these values in order to obtain potential temperatures that are applied in the Boussinesq geodynamic model (see Fig. 2). 50%-solidus temperatures are simply calculated as the average between the solidus and liquidus temperatures.

Once the cumulate layer starts to crystallize, thermal and compositional density variations control solid-state flow. Density $\rho_{solid}$ is a function of cumulate temperature and iron content: $\rho_{solid} = \rho_0 + \alpha \cdot T - \alpha \cdot T_0 + C \cdot Fe\#_{solid}$ with temperature $T$, reference temperature $T_0 = T_{CMB}$, reference density $\rho_0$, $C$ = 850 kg/m$^3$, and $Fe\#_{solid} = Fe_{solid}/(Mg_{solid} + Fe_{solid})$. The viscosity of the cumulate layer (and after freezing is complete: of the mantle) $\eta_{mantle}$ is assumed to be constant. It is varied as a free parameter between $1.17 \cdot 10^{20}$ Pa·s and $4.69 \cdot 10^{20}$ Pa·s (see Table 2). These viscosities translate into global (thermal) Rayleigh numbers $Ra_T$ of $1.15 \cdot 10^8$ to $4.6 \cdot 10^8$.

**Table 1:** Key model parameters. Parameters that have been varied are given in bold (see Table 2 for parameter variations).

| **Parameter** | **symbol** | **value** |
|---|---|---|
| depth of the numerical domain | $z_{box}$ | 2,900 km |
| width of the numerical domain | $x_{box}$ | 2,900 km |
| surface temperature | $T_0$ | 0 ºC |
| mantle reference density | $\rho_0$ | 4,000 kg/m$^3$ |
| thermal expansivity | $\alpha$ | $1.5 \cdot 10^{-5}$ K$^{-1}$ |
| gravity acceleration | $g$ | 9.8 m$^2$/s |
| thermal diffusivity | $\kappa$ | $0.8 \cdot 10^{-6}$ m$^2$/s |
| **initial depth extent of magma ocean** | $\boldsymbol{d_{MO,init}}$ | **see text and Table 2** |
| **progression velocity of MO crystallization front** | $\boldsymbol{v_{MOC}}$ | **see text and Table 2** |
| **mantle reference viscosity** | $\boldsymbol{\eta_{mantle}}$ | **see text and Table 2** |
| **mantle reference potential temperature** | $\boldsymbol{T_{CMB}}$ | **see text and Table 2** |

**Table 2**: Summary of all cases. Model parameters are given in columns two to six. Model predictions are reported in the two columns to the right in bold italics. $V_{unmixed}$ and $\Delta\rho_{comp}$ are the volume and average density anomaly (i.e., relative to the rest of the mantle) of the unmixed layer near the CMB at 3 Gyrs model time. For description of model parameters, see Table 1. Model predictions are also visualized in Figure 6. Initial conditions are visualized in Figure 2 as a function of freezing scenario (column two). (*) $T_{CMB}$ is 3000 °C for all cases in the solidus thermal scenario, and 3500 °C for all other cases (see column three).

| case | freezing scenario | thermal scenario* | $\eta_{mantle}$ [Pa·s] | $v_{MOC}$ [cm/yr] | $d_{MO,init}$ [km] | $V_{unmixed}$ [%] | $\Delta\rho_{comp}$ [kg/m$^3$] |
|---|---|---|---|---|---|---|---|
| **F_LRa1** | F | solidus | $4.69 \cdot 10^{20}$ | 0.29 | 2610 | ***7.3*** | ***155.9*** |
| **F_LRa2** | F | solidus | $4.69 \cdot 10^{20}$ | 0.58 | 2610 | ***7.2*** | ***185.0*** |
| **F_LRa3** | F | solidus | $4.69 \cdot 10^{20}$ | 1.45 | 2610 | ***11.6*** | ***163.2*** |
| **F_LRa4** | F | solidus | $4.69 \cdot 10^{20}$ | 2.9 | 2610 | ***13.7*** | ***176.3*** |
| **F_LRa5** | F | solidus | $4.69 \cdot 10^{20}$ | 5.8 | 2610 | ***20.0*** | ***169.3*** |
| **F_LRa6** | F | solidus | $4.69 \cdot 10^{20}$ | 14.5 | 2610 | ***23.8*** | ***161.6*** |
| **F_LRa7** | F | solidus | $4.69 \cdot 10^{20}$ | 29 | 2610 | ***27.6*** | ***173.0*** |
| **F_LRa8** | F | solidus | $4.69 \cdot 10^{20}$ | 58 | 2610 | ***24.4*** | ***174.1*** |
| **F_MRa1** | F | solidus | $2.34 \cdot 10^{20}$ | 0.29 | 2610 | ***5.6*** | ***150.0*** |
| **F_MRa2** | F | solidus | $2.34 \cdot 10^{20}$ | 0.58 | 2610 | ***6.8*** | ***149.4*** |
| **F_MRa3** | F | solidus | $2.34 \cdot 10^{20}$ | 1.45 | 2610 | ***8.3*** | ***188.1*** |
| **F_MRa4** | F | solidus | $2.34 \cdot 10^{20}$ | 2.9 | 2610 | ***11.6*** | ***164.8*** |
| **F_MRa5** | F | solidus | $2.34 \cdot 10^{20}$ | 5.8 | 2610 | ***13.8*** | ***189.5*** |
| **F_MRa6** | F | solidus | $2.34 \cdot 10^{20}$ | 14.5 | 2610 | ***15.9*** | ***193.1*** |
| **F_MRa7** | F | solidus | $2.34 \cdot 10^{20}$ | 29 | 2610 | ***23.5*** | ***162.3*** |
| **F_MRa8** | F | solidus | $2.34 \cdot 10^{20}$ | 58 | 2610 | ***27.7*** | ***172.2*** |
| **F_HRa1** | F | solidus | $1.17 \cdot 10^{20}$ | 0.29 | 2610 | ***4.0*** | ***171.0*** |
| **F_HRa2** | F | solidus | $1.17 \cdot 10^{20}$ | 0.58 | 2610 | ***5.6*** | ***168.7*** |
| **F_HRa3** | F | solidus | $1.17 \cdot 10^{20}$ | 1.45 | 2610 | ***7.5*** | ***186.0*** |
| **F_HRa4** | F | solidus | $1.17 \cdot 10^{20}$ | 2.9 | 2610 | ***8.5*** | ***196.2*** |
| **F_HRa5** | F | solidus | $1.17 \cdot 10^{20}$ | 5.8 | 2610 | ***11.3*** | ***173.8*** |
| **F_HRa6** | F | solidus | $1.17 \cdot 10^{20}$ | 14.5 | 2610 | ***15.6*** | ***165.9*** |
| **F_HRa7** | F | solidus | $1.17 \cdot 10^{20}$ | 29 | 2610 | ***14.4*** | ***201.6*** |
| **F_HRa8** | F | solidus | $1.17 \cdot 10^{20}$ | 58 | 2610 | ***23.3*** | ***162.2*** |
| **I_LRa** | I | solidus | $4.69 \cdot 10^{20}$ | - | 0 | ***19.7*** | ***108.2*** |
| **I_MRa** | I | solidus | $2.34 \cdot 10^{20}$ | - | 0 | ***20.3*** | ***101.8*** |
| **I_HRa** | I | solidus | $1.17 \cdot 10^{20}$ | - | 0 | ***21.5*** | ***95.8*** |
| **EI_LRa** | $E_{1450}I$ | solidus | $4.69 \cdot 10^{20}$ | - | 0 | ***6.2*** | ***127.0*** |
| **EI_MRa** | $E_{1450}I$ | solidus | $2.34 \cdot 10^{20}$ | - | 0 | ***5.9*** | ***128.9*** |
| **EI_HRa** | $E_{1450}I$ | solidus | $1.17\cdot10^{20}$ | - | 0 | ***5.1*** | ***136.8*** |
| **IF_HRa1** | $I_{290}F$ | solidus | $1.17\cdot10^{20}$ | 0.29 | 290 | ***0*** | ***-*** |
| **IF_HRa2** | $I_{290}F$ | solidus | $1.17\cdot10^{20}$ | 0.58 | 290 | ***0.9*** | ***223.5*** |
| **IF_HRa3** | $I_{290}F$ | solidus | $1.17\cdot10^{20}$ | 1.45 | 290 | ***1.4*** | ***298.2*** |
| **IF_HRa4** | $I_{290}F$ | solidus | $1.17\cdot10^{20}$ | 2.9 | 290 | ***2.2*** | ***246.8*** |
| **IF_HRa5** | $I_{290}F$ | solidus | $1.17\cdot10^{20}$ | 5.8 | 290 | ***3.7*** | ***220.5*** |
| **IF_HRa6** | $I_{290}F$ | solidus | $1.17\cdot10^{20}$ | 14.5 | 290 | ***5.9*** | ***187.5*** |
| **IF_HRa7** | $I_{290}F$ | solidus | $1.17\cdot10^{20}$ | 29 | 290 | ***5.3*** | ***247.0*** |
| **IF_HRa8** | $I_{290}F$ | solidus | $1.17\cdot10^{20}$ | 58 | 290 | ***5.3*** | ***270.7*** |
| **EIF_HRa1** | $E_{1450}I_{290}F$ | solidus | $1.17\cdot10^{20}$ | 0.29 | 290 | ***1.5*** | ***261.0*** |
| **EIF_HRa2** | $E_{1450}I_{290}F$ | solidus | $1.17\cdot10^{20}$ | 0.58 | 290 | ***1.5*** | ***273.5*** |
| **EIF_HRa3** | $E_{1450}I_{290}F$ | solidus | $1.17\cdot10^{20}$ | 1.45 | 290 | ***2.0*** | ***233.8*** |
| **EIF_HRa4** | $E_{1450}I_{290}F$ | solidus | $1.17\cdot10^{20}$ | 2.9 | 290 | ***2.3*** | ***243.7*** |
| **EIF_HRa5** | $E_{1450}I_{290}F$ | solidus | $1.17\cdot10^{20}$ | 5.8 | 290 | ***3.0*** | ***264.9*** |
| **EIF_HRa6** | $E_{1450}I_{290}F$ | solidus | $1.17\cdot10^{20}$ | 14.5 | 290 | ***3.9*** | ***223.0*** |
| **EIF_HRa7** | $E_{1450}I_{290}F$ | solidus | $1.17\cdot10^{20}$ | 29 | 290 | ***3.8*** | ***249.5*** |
| **EIF_HRa8** | $E_{1450}I_{290}F$ | solidus | $1.17\cdot10^{20}$ | 58 | 290 | ***3.7*** | ***250.8*** |
| **F^HRa1** | F | liquidus | $1.17\cdot10^{20}$ | 0.29 | 2610 | ***3.4*** | ***197.5*** |
| **F^HRa2** | F | liquidus | $1.17\cdot10^{20}$ | 0.58 | 2610 | ***3.8*** | ***199.3*** |
| **F^HRa3** | F | liquidus | $1.17\cdot10^{20}$ | 1.45 | 2610 | ***5.2*** | ***219.3*** |
| **F^HRa4** | F | liquidus | $1.17\cdot10^{20}$ | 2.9 | 2610 | ***5.2*** | ***220.7*** |
| **F^HRa5** | F | liquidus | $1.17\cdot10^{20}$ | 5.8 | 2610 | ***6.8*** | ***225.2*** |
| **F^HRa6** | F | liquidus | $1.17\cdot10^{20}$ | 14.5 | 2610 | ***10.0*** | ***209.7*** |
| **F^HRa7** | F | liquidus | $1.17\cdot10^{20}$ | 29 | 2610 | ***16.2*** | ***199.3*** |
| **F^HRa8** | F | liquidus | $1.17\cdot10^{20}$ | 58 | 2610 | ***15.4*** | ***191.5*** |
| **I^HRa** | I | 50%-sol. | $1.17\cdot10^{20}$ | - | 0 | ***17.6*** | ***106.0*** |
| **EI^HRa** | $E_{1450}I$ | 50%-sol. | $1.17\cdot10^{20}$ | - | 0 | ***3.1*** | ***162.0*** |
| **IF^HRa1** | $I_{290}F$ | 50%-sol. | $1.17\cdot10^{20}$ | 0.29 | 290 | ***0*** | ***-*** |
| **IF^HRa2** | $I_{290}F$ | 50%-sol. | $1.17\cdot10^{20}$ | 0.58 | 290 | ***0*** | ***-*** |
| **IF^HRa3** | $I_{290}F$ | 50%-sol. | $1.17\cdot10^{20}$ | 1.45 | 290 | ***0*** | ***-*** |
| **IF^HRa4** | $I_{290}F$ | 50%-sol. | $1.17\cdot10^{20}$ | 2.9 | 290 | ***2.2*** | ***249.5*** |
| **IF^HRa5** | $I_{290}F$ | 50%-sol. | $1.17\cdot10^{20}$ | 5.8 | 290 | ***2.7*** | ***257.5*** |
| **IF^HRa6** | $I_{290}F$ | 50%-sol. | $1.17\cdot10^{20}$ | 14.5 | 290 | ***4.3*** | ***229.8*** |
| **IF^HRa7** | $I_{290}F$ | 50%-sol. | $1.17\cdot10^{20}$ | 29 | 290 | ***4.7*** | ***267.0*** |

## 3. Magma-ocean freezing and cumulate fractionation

The initial compositional profiles for mantle convection are shaped by fractionation (melt-solid separation) and mixing (convection) processes during MO freezing. The balance between these processes is controlled by the rate of MO cooling. In particular, MO cooling rate determines the progression rate (or velocity) of the crystallization front $v_{MOC}$, with which fractionation processes such as crystal sinking and melt-solid separation compete. For vigorous convection in the MO, which leads to efficient cooling and thus high $v_{MOC}$, any growing crystals tend to remain entrained in the MO such that little or no compositional fractionation occurs (equilibrium crystallization, i.e. equivalent to batch crystallization). In the extreme end-member case, a thick mushy layer grows readily near the base of the MO, with $v_{MOC}$ being faster than melt-solid separation in the mush layer (*Tonks and Melosh*, 1990). In contrast, for sluggish convection in the MO and thus low $v_{MOC}$, any growing crystals may settle, or any mushy layers may undergo effective melt-solid segregation, such that strong (due to fractional crystallization) or at least moderate compositional fractionation occurs.

Accordingly, we consider three modes of crystallization. The fast-freezing end-member with very vigorous convection in the liquid MO is trivial. The timescale of MO crystallization $\tau_{MOC} = z_{box} / v_{MOC}$ is significantly shorter than that of crystal settling in the MO $\tau_{CS}$ or melt-solid separation in the mushy layer $\tau_{MSS}$. Thus, no compositional fractionation occurs (equilibrium crystallization), and the MO is virtually quenched. In this scenario (E), a fully homogeneous mantle composition is implied (not explicitly modeled).

For (F) fractional crystallization, $\tau_{MOC}$ is instead much larger than both $\tau_{CS}$ and $\tau_{MSS}$. Thus, crystals settle to form orthocumulate (i.e., incipient cumulate) layer(s) at the base of the MO, and these orthocumulate layer(s) undergo efficient compaction. The composition of the resulting cumulate composition is very Fe-poor, being defined by melt-solid equilibria at (infinitesimally) small degrees of crystallization. Hence, the remaining MO evolves progressively towards Fe-rich compositions, and so do the coexisting cumulate layers (Figure 2a). Even though we only consider the fractionation of Fe and Mg using a constant $K_D$, we find compositional density profiles that are similar to those in the more realistic mineralogical models of *Elkins-Tanton* (2008).

In an (I) intermediate case, $\tau_{MOC}$ is similar to timescales $\tau_{CS}$ and $\tau_{MSS}$. For example, (Ia) $\tau_{MSS} < \tau_{MOC} < \tau_{CS}$, or (Ib) $\tau_{MSS} > \tau_{MOC} > \tau_{CS}$. Note that under the extreme conditions of the MO, it is unclear whether $\tau_{MSS} < \tau_{CS}$, or $\tau_{MSS} > \tau_{CS}$. In the former case (Ia), crystals remain entrained in the MO, but any mushy layers at the base of the MO should undergo efficient melt-solid segregation as soon as the rheological transition at crystal fractions of ~60% is reached (*Abe*, 1995; *Arzi*, 1978; *Costa et al.*, 2009). In the latter case (Ib), crystals sink to the base of the MO, but efficient compaction of the orthocumulate layer (that consists of these crystals and interstitial MO liquid) does not occur on the relevant timescales. Cumulate compositions in (Ia) and (Ib) are calculated from melt-solid equilibria at 60% crystallization, and by adding 40% interstitial non-fractionated MO liquid to low-degree fractionally-crystallized solid, respectively. Figure 2a shows that the resulting compositional profiles for both these variants of intermediate MO fractionation are less steep than that for (F) fractional crystallization, and very similar to each other. Hereinafter, we consider variant (Ia) as an intermediate mode (I) of MO crystallization, keeping in mind that it is representative of various physical processes that may lead to an intermediate efficiency of chemical fractionation.

The specific relevance of the above three end-member modes (E, I, F) of crystallization depends on the cooling history of the MO. In realistic MO evolution models, $v_{MOC}$ steadily decreases during the progression of MO freezing. Figure 3 shows that the heat flux out of the MO and into space, and thus $v_{MOC}$, systematically decreases with decreasing MO depth due to progressive upward crystallization (*Lebrun et al.*, 2013; *Solomatov et al.*, 1993). This effect is primarily explained by MO surface temperatures being tied to the relevant depth-dependent crystallization temperatures. In addition, exhalation of a proto atmosphere slows down MO cooling. During the early stages of MO crystallization, little or no atmosphere may have been present to shield radiation out of the MO (cf. *Genda and Abe*, 2003). During the late stages, a progressively thick steam atmosphere is exhaled from the MO to significantly slow down MO cooling (*Elkins-Tanton*, 2008; *Hamano et al.*, 2013; *Lebrun et al.*, 2013; *Zahnle et al.*, 2007).

While parameter uncertainties in scaling relationships do not allow to exclude even the end-member scenarios of MO crystallization (*Solomatov*, 2015), any realistic evolutionary path should proceed from (fast-freezing) non-fractionating toward (slow-freezing) fractionating scenarios, and not vice-versa. Keeping large uncertainties in mind, the transitional heat-flux, below which fractional crystallization becomes dominant, has been roughly estimated at $10^4$~$10^6$ W/m$^2$ (*Solomatov*, 2015). Therefore, it appears likely that the style of MO crystallization has switched from equilibrium to intermediate, and/or from intermediate to fractional crystallization somewhere in the upper mantle (see background shading in Fig. 3). While a dominance of fractional crystallization through the entire mantle cannot be excluded, the dominance of equilibrium crystallization in the uppermost mantle remains unlikely, because unhindered grain-growth due to Ostwald ripening should occur once the entire upper mantle drops below the liquidus (*Solomatov,* 2015). Also note that the partially molten mantle beneath mid-ocean ridges, i.e. the best modern analog for a late-stage MO, displays porosities of only 1%-2% (*Forsyth et al.*, 1998).

Along these lines, we explore five scenarios of MO crystallization. In scenarios (F) and (I), fractional and intermediate crystallization occur over the full mantle depth range, respectively. In scenario ($I_{290}F$), or in short (IF), intermediate crystallization occurs at depths >290 km, and then switches to fractional crystallization in the uppermost mantle (see above). In scenario ($E_{1450}I$), or (EI), equilibrium crystallization is dominant at depths >1450 km, and then switches to intermediate crystallization. Finally, in scenario ($E_{1450}I_{290}F$), or (EIF), equilibrium crystallization switches to intermediate crystallization at 1450 km depth, and then further switches to fractional crystallization at 290 km depth. No chemical fractionation in the deep mantle (>1450 km) may be related to a mid-mantle initial depth extent of the MO (*Wood et al.*, 2006), or extremely fast freezing of the very deep and hot MO. For initial compositional and thermal profiles, see Figures 2b-2c. According to the above discussion of MO cooling models, we consider scenarios (F), (IF) and (EIF) more likely that scenarios (I) and (EI), or even than end-member (E) equilibrium crystallization (not explored here).

## 4. Cumulate convection and mantle mixing

In this section, we describe the results of our geodynamic simulations for crystallization scenarios (F), (I), (IF), (EI), and (EIF). In these simulations, we model convection and mixing within the growing cumulate layers during MO freezing (except for fast-freezing scenarios (I) and (EI), see below), and throughout the mantle after MO freezing is complete. Model predictions in terms of compositional mantle heterogeneity after billions of years of solid-state convection (i.e., ~3 Gyrs after the initiation of MO freezing) are compared to the structure of the modern mantle. We terminate our models after 3 Gyrs model time to limit the amount of model timesteps (i.e., usually on the order of $10^5$), and any artificial mantle mixing due to numerical diffusion.

### *4.1 Fractional Crystallization of the MO*

In the slow-freezing fractional-crystallization scenario (F), we explicitly solve for solid-state convection during the progression of the crystallization front (i.e., $d_{MO,init}$ = 2610 km). As this progression is relatively slow and comparable to convective timescales (see above), incremental convective overturns may occur within the growing cumulate layer to perturb the progressive compositional stratification that results from MO freezing (Fig. 2). We find that solid-state convection initiates within the layer as soon as the cumulate layer reaches a critical thickness. Convection is driven by unstable compositional stratification due to progressive upward Fe-enrichment (Fig. 2b), as well as unstable thermal stratification (Fig. 2c). Compositional stratification is dominant in driving the flow, particularly during the late stages of MO freezing. Convection is almost exclusively top-down driven as an unstable thermochemical boundary layer is constantly re-established at the moving top boundary due to progressive MO freezing; continuous cool downwellings inhibit the growth of a hot thermal boundary layer at the CMB.

The wavelength of convection in the cumulate layer (or number of downwellings) is controlled by the viscosity of the cumulate $\eta_{mantle}$ and the progression velocity of the MO crystallization front $v_{MOC}$. Parameters $\eta_{mantle}$ and $v_{MOC}$ are independent of each other: $\eta_{mantle}$ is the viscosity of the solid; $v_{MOC}$ is related to the surface heat flux out of the MO (see above and Fig. 3). With increasing $\eta_{mantle}$, the wavelength of convection within the cumulate increases, and numbers of downwellings decrease. Increasing $v_{MOC}$ has the same effect, because it increases the layer thickness that is progressively added to the "backside" of the boundary layer during growth of convective instability to sustain long-wavelength convection (see Figure 4, top two rows).

As the mixing efficiency scales with the wavelength of convection (or the number of downwellings that sink trough the cumulate layers), parameters $\eta_{mantle}$ and $v_{MOC}$ control the initial thermochemical structure of the mantle. Figure 5 shows compositional profiles just after MO crystallization is complete. Much of the shallow proto-mantle is generally well mixed. This material has originally (i.e., before overturn) crystallized over a range of depths, and progressively been perturbed by subsequent downwellings. Only the deepest layers of the proto-mantle remain poorly mixed after overturn. This dense, strongly Fe-enriched material has originally crystallized at shallow depths during the late stages of MO freezing, and remained unperturbed by any subsequent downwellings. The thickness of this deep dense layer is related to the boundary layer thicknesses $d_{crit}$ of the final incremental overturn, which in turn is controlled by $\eta_{mantle}$ and $v_{MOC}$ (Fig. 4).

We find a linear trade-off between $\eta_{mantle}$ and $v_{MOC}$, both in terms of the wavelengths of convection during cumulate overturn, as well as the compositional profile after cumulate overturn (Appendix A, Fig. A1). Decreasing $v_{MOC}$ by a factor of two compensates for a factor-of-two increase in $\eta_{mantle}$. This linear trade-off, which is confirmed experimentally (Fig. A1) and theoretically (Appendix A) allows us to significantly reduce computational cost by exploring parameter ranges of $\eta_{mantle}$ and $v_{MOC}$ that are higher and smaller, respectively (i.e., by 2-3 orders of magnitude), than those inferred for the early Earth.

Stable stratification after overturn is difficult to be removed by subsequent solid-state mantle convection. The deep dense layer is stable, because it is more Fe-enriched and cooler that the rest of the proto-mantle. It is cool at first, because it has originally crystallized at low freezing temperatures near the surface, and thermal equilibration is predicted to occur on timescales of 100s of Myrs. Note however that these predicted timescales are clearly overestimates. For realistic parameters of $\eta_{mantle}$ and $v_{MOC}$, downwellings should be equilibrated before they actually reach the lowermost mantle; see Appendix A2. In any case, the deep dense layer is predicted to remain stable beyond thermal equilibration; negative compositional buoyancy is sufficient for stabilization. Persistent separation of the mantle into two, or more, layers persists as a critical buoyancy number $B = \Delta\rho_{comp} / \rho_0 \alpha T_{CMB}$ (with $\Delta\rho_{comp}$ the density difference between layers) of ~0.3 is exceeded (*Li et al.*, 2014b; *Tosi et al.*, 2013). These $B$ are exceeded in all our models with fractional crystallization.

The volume $V_{unmixed}$ of the deep dense layer after 3 Gyrs model time hence depends on the compositional profile just after overturn (Fig. 5), which controls $B$. Accordingly, $V_{unmixed}$ is ultimately controlled by parameters $\eta_{mantle}$ and $v_{MOC}$, predicted to decrease with decreasing $\eta_{mantle}$ or $v_{MOC}$ (see Fig. 6). This predicted decrease is well explained by the effects of $\eta_{mantle}$ and $v_{MOC}$ on cumulate mixing during MO freezing. This finding demonstrates that MO-freezing processes can impact subsequent solid-state mantle dynamics, as well as the present-day make-up of the mantle.

Assumptions on the thermal scenario (upper bound (liquidus) vs. lower-bound (solidus) thermal scenarios) also affect $V_{unmixed}$ and $\Delta\rho_{comp}$. In the upper-bound thermal case(s), $B$ is smaller than in the lower-bound thermal case(s), because $T_{CMB}$ is larger (by 1000 K), causing a systematic decrease in $V_{unmixed}$ (Fig. 6a vs. Fig. 6b). This decrease is associated with an increase in $\Delta\rho_{comp}$ due to a concentration effect of the most Fe-enriched material. Our models with high $T_{CMB}$ further explicitly predict that the solidus in the deep dense layer (i.e., in the bottom 200-300 km of the mantle) is exceeded after thermal equilibration. This prediction implies the formation of a secondary basal magma ocean (BMO), independent of the possible formation of a (primary) BMO during MO crystallization (Labrosse et al., 2007; see discussion section). $T_{CMB}$ should be generally higher (i.e., closer to the liquidus of mantle rocks) in fractional-crystallization scenarios than other modes of MO crystallization. Thus, the mode of crystallization in the deep mantle may control the thermal (through initial $T_{CMB}$) (*Andrault et al.*, 2016) and chemical (through fractionation of a large BMO) evolution of our planet.

### *4.2 Equilibrium-crystallization and intermediate-crystallization scenarios*

In the equilibrium and intermediate-crystallization scenarios (E), (I) and (EI), we ignore cumulate overturn during MO crystallization. Accordingly, the MO is already completely crystallized at the start of the simulations (i.e., $d_{MO,init}$ = 0 km), because the timescale for Rayleigh-Taylor instability $\tau_{RT}$ is assumed to exceed that of the progression of the crystallization front $\tau_{MOC}$ in this relatively fast-freezing scenario (i.e., $\tau_{MOC}$ is effectively zero). Thus, one large big overturn occurs shortly after the start of the simulation (Figs. 7a, 7d). As no incremental overturns occur, the compositional stratification just after this overturn, as well as the final stratification after ~3 Gyrs (as represented by $V_{unmixed}$ and $\Delta\rho_{comp}$) is virtually independent of $\eta_{mantle}$ (see Fig. 6). The final compositional stratification depends almost exclusively on the crystallization and thermal scenarios. $V_{unmixed}$ is zero for end-member equilibrium crystallization (not modeled), and non-zero otherwise. $V_{unmixed}$ is smaller for scenario (EI) than for scenario (I), because the total volume of strongly Fe-enriched material is smaller in the initial profile (Fig. 2a). $V_{unmixed}$ also tends to be smaller (and $\Delta\rho_{comp}$ to be higher) for the upper-bound thermal scenario than for the lower-bound thermal scenario due to higher $T_{CMB}$ (for explanation, see section 4.1). Overall, the trend defined by intermediate-crystallization cases (I) and (EI) is shifted towards lower $\Delta\rho_{comp}$ at a given $V_{unmixed}$ relative to fractional-crystallization cases (Fig. 6). This result is explained by the initial compositional profile (Fig. 2b) being significantly less steep near the top for (I)/(EI) than for (F), and initial $Fe\#_{solid}$ being smaller at a given depth.

### *4.3 Combined Crystallization Scenarios*

In the combined scenarios (IF) and (EIF), we assume that the MO freezes by intermediate (IF) or equilibrium (EIF) crystallization in the deep mantle at depths >1450 km, by intermediate crystallization in the mid mantle, and by fractional crystallization in the shallow mantle at depths <290 km (Figure 2b). Such an evolution may be related to the progressive deceleration of MO freezing, or a limited initial depth extent of the MO (see section 3). In (IF)/(EIF), we only explicitly solve for cumulate convection during the slow freezing of the last 10% of MO cumulates at depths smaller than $d_{MO,init}$ = 290 km (i.e., $\tau_{MOC}$ = 290 km / $v_{MOC}$.). Accordingly, one major overturn happens after 90% MO freezing (Figs. 7g, 7j) with only a few incremental overturns thereafter (Figs. 7h. 7k).

We find that in all cases in scenarios (IF) and (EIF), $V_{unmixed}$ is generally small (<4%), and $\Delta\rho_{comp}$ is large (~6%); see Figure 6. $V_{unmixed}$ is small, because strong iron enrichment in MO cumulates is limited to a small initial depth range very close to the surface, and the initial compositional profile is relatively flat elsewhere (Fig. 2b). In turn, $\Delta\rho_{comp}$ is large because the initial compositional profile is very steep near the top (i.e., steeper than for fractional-crystallization cases). The differences in terms of $V_{unmixed}$ and $\Delta\rho_{comp}$ between various cases in scenarios (IF) and (EIF) remain rather small, because compositional profiles are similar, and the sequence of overturn(s) (see above) is controlled by the choice of $d_{MO,init}$. Accordingly, the effects of $v_{MOC}$ on cumulate mixing are small. Similarly, general model predictions are virtually independent of the thermal scenario assumed (i.e. of $T_{CMB}$).

## 5. Comparison of model predictions with present-day mantle structure

A key prediction of our models is that incremental cumulate overturns results in a stable Fe-rich layer at the base of the mantle that remains largely unmixed by subsequent solid-state mantle convection over billions of years. The volume $V_{unmixed}$ and density anomaly $\Delta\rho_{comp}$ of this unmixed layer mostly depend on the mode of crystallization, as well as parameters $\eta_{mantle}$ and $v_{MOC}$ within the subset of fractional-crystallization cases (see section 4). As the mode of crystallization itself is controlled by $v_{MOC}$ (see section 3), the speed $v_{MOC}$ or corresponding timescale $\tau_{MOC}$ of MO freezing is arguably the key parameter that determines the initial condition of mantle convection, as well as subsequent evolution. We therefore motivate more detailed studies of coupled MO and early atmospheric evolution.

Figure 6 shows trends of $\Delta\rho_{comp}$ as a function of $V_{unmixed}$ for all our cases. Fractional crystallization scenarios (F) yield moderate $\Delta\rho_{comp}$ (about 4%-5%) at any given $V_{unmixed}$. In turn, equilibrium and intermediate MO crystallization scenarios ((I) and (EI)) imply low $\Delta\rho_{comp}$ (<3%). Finally, combined crystallization scenarios ((IF) and (EIF)) yield high $\Delta\rho_{comp}$ and small $V_{unmixed}$. These general trends are independent of the thermal scenario assumed (Fig. 6a vs. Fig. 6b), a finding that is explained the dominance of compositional instability over thermal instability during initial (incremental) overturn(s) (see Figure 2).

### *5.1 Large low shear-wave velocity provinces as a constraint for MO crystallization*

Model predictions in terms of the volume $V_{unmixed}$ and density anomaly $\Delta\rho_{comp}$ of the deep unmixed layer can be compared to the present-day make-up of the Earth's mantle. The mantle is thought to be largely homogeneous (and pyrolitic) to first order, except for the large low shear-wave velocity provinces (LLSVPs) at the base of the mantle (*Garnero and McNamara*, 2008), which are interpreted as intrinsically dense thermochemical piles (*Davaille*, 1999; *McNamara and Zhong*, 2004). LLSVP density anomalies $\Delta\rho_{LLSVP}$ can be bracketed on the basis of LLSVP shapes with 90 kg/m$^3$ ≤ $\Delta\rho_{LLSVP}$ ≤ 150 kg/m$^3$ (*Deschamps and Tackley*, 2009). The nature of the LLSVPs remains under debate (*Brandenburg et al.*, 2008; *Davies et al.*, 2012; *Tackley*, 2012), but an ancient origin, perhaps related to MO cumulates, has been put forward (*Ballmer et al.*, 2016; *Bian et al.*, 2016; *Deschamps et al.*, 2012; *Li et al.*, 2014a; *Tackley and Xie*, 2002). Iron and silica enriched LLSVP compositions are consistent with seismic observations (*Deschamps et al.*, 2012; *Trampert et al.*, 2004). An ancient origin of LLSVPs is supported by noble-gas isotopic signatures of hotspot lavas (*Mukhopadhyay*, 2012; *Williams et al.*, 2015), which are fed by plumes from LLSVP tops and/or margins (*Austermann et al.*, 2014; *Davies et al.*, 2015; *Torsvik et al.*, 2006; *Weis et al.*, 2011). The LLSVPs are inferred to make up a volume $V_{LLSVP}$ ≈ 3% of that of the mantle (*Hernlund and Houser*, 2008). Recent estimates of $V_{LLSVP}$ ≈ 7% are based on the robust detection of coherent slow material in the mid-mantle (*Cottaar and Lekic*, 2016). This estimate, however, likely includes significant amounts of non-primordial material (i.e., unrelated to MO freezing) such as hot pyrolite, and/or subducted basalt (*Ballmer et al.*, 2016). Therefore, we consider the conservative estimate of $V_{LLSVP}$ ≈ 3% as an upper-bound for the volume of intrinsically dense primordial material that persists in the deep mantle. Our approach is to use LLSVP properties as a constraint for early-Earth processes, such as MO freezing.

To distinguish between MO crystallization scenarios, we compare model predictions in terms of $V_{unmixed}$ and $\Delta\rho_{comp}$ with estimates for LLSVP volumes and density anomalies (golden shading in Fig. 6). For example, we can rule out scenarios with (I) intermediate crystallization throughout the mantle, or fractional-crystallization scenarios with high $v_{MOC}$. In contrast, cases (EI) with a switch from equilibrium to intermediate crystallization in the mid-mantle, or fractional-crystallization scenarios with low $v_{MOC}$ predict $V_{unmixed}$ and $\Delta\rho_{comp}$ that are similar to LLSVP properties estimated for the present-day. Note however that the fractional-crystallization cases with low $v_{MOC}$ imply $\tau_{MOC}$ of 100s Myrs, i.e. much higher than those estimated for the maximum lifetime of the MO, for example based on the age of the oldest detrital zircons (4.38 Ga) (*Harrison*, 2009; *Mojzsis et al.*, 2001; *Wilde et al.*, 2001). In turn, the modeled $\eta_{mantle}$ (~$10^{20}$ Pa·s) are are also much higher than realistic for the near-solidus proto mantle. Due to the experimentally (Fig. A1) and theoretically (Appendix A1) confirmed linear trade-off between parameters $v_{MOC}$ and $\eta_{mantle}$, we are confident that our results can be extrapolated to the relevant conditions, which may be 2-3 orders of magnitude greater than modeled in terms of $v_{MOC}$ and $\eta_{mantle}$ (or smaller in terms of $\tau_{MOC}$). As we assume that $v_{MOC}$ and $\eta_{mantle}$ are independent of time and of temperature (and/or depth and/or stress), respectively, the applied (or extrapolated) $v_{MOC}$ and $\eta_{mantle}$ are relevant for the uppermost mantle, where the relevant late-stage cumulate overturn(s) occur (also see Fig. 3b).

Beyond these limitations due to our model assumptions, our models remain systematically restricted as important geological processes during the secular evolution of the mantle are not explicitly considered. For example, the present-day LLSVPs may consist of a mix of MO cumulates and basaltic material that has been subducted over Earth's history and piled up in the lowermost mantle (*Tackley*, 2012). Any addition of basaltic material to the deep unmixed layer would tend to increase $V_{unmixed}$ and decrease $\Delta\rho_{comp}$. Also, entrainment of the deep dense layer may be enhanced as internal heating is considered, particularly as radioactive elements are partitioned into the layer (*Brown et al.*, 2014). In turn, entrainment and mixing across the whole mantle may be delayed due to the (depth-dependent) effects of composition on density and rheology (*Ballmer et al.*, 2015; 2017; *Manga*, 2010; *Nakagawa and Buffett*, 2005). Finally, progressive fractionation of the deep dense layer may occur as it evolves through a basal-magma-ocean stage (see discussion). Keeping in mind these additional processes, we consider somewhat looser constraints in terms of $V_{unmixed}$ and $\Delta\rho_{comp}$ (light golden shading in Fig. 6) than are provided by the present-day properties of LLSVPs.

### *5.2 Stable stratification at the base of the mantle*

Even considering these looser constraints, all our models have difficulties to reconcile the present-day structure of the Earth's mantle. All our models predict that the deep unmixed layer is compositionally stratified internally, collapsing into different sub-layers that do not mix with each other. Note that $\Delta\rho_{comp}$ as evaluated above (Fig. 6) is merely an average density anomaly, integrated over $V_{unmixed}$. Figure 8 shows that Fe-contents and intrinsic density anomalies progressively increase downward through the deep unmixed layer, even after 3 Gyr model time. Internal stable stratification is somewhat less intense in crystallization scenarios (I) and (EI) than in scenarios (F), (IF) and (EIF), but note that the latter cases are more realistic than the former as they evolve through a stage of fractional crystallization at least in the late-stage uppermost-mantle MO (see section 3 and Fig. 3a).

This robust result is inconsistent with the structure of the present-day mantle. For example, the most enriched sub-layer at the base of the mantle (originally formed during the final stages of MO crystallization close to the surface) is predicted to form a thin (10s of km) global layer with density anomalies of ~10% or greater. This sub-layer makes up 0.1%~1% of the Earth's mantle and involves $Fe\#_{solid} > 0.4$ (Figures 5, 8). Most importantly, the above Fe-contents and densities are much higher than those inferred for the present-day LLSVPs (i.e., 2-3% (*Deschamps and Tackley*, 2009) and $Fe\#_{solid} < 0.14$ (*Deschamps et al.*, 2012), respectively). Also note that the seismically-imaged Ultra-Low Velocity Zones (e.g., *Garnero et al.*, 1998) are too small (i.e., ~0.01% of that of the mantle (*Dobson and Brodholt*, 2005; *Persh and Vidale*, 2004)) to host this very Fe-rich material. Thus, our models appear to neglect an important process that can successively reduce the Fe-content of the final-stage MO cumulates.

### *5.3 Hybridization during Incremental Overturns*

In order to identify this process, we estimate the critical boundary-layer thickness during the final stages of MO cooling, based on our estimate for $v_{MOC}$ as a function of $d_{MO}$ (Fig. 3b). This critical boundary-layer thickness scales with the wavelength of convective instability during incremental overturns (see Appendix A2), and thus the size of related cumulate diapirs that sink through the proto-mantle. Figure 9a shows the predicted size of cumulate diapirs as a function of MO depth extent (i.e. decreasing with time) for two different MO cooling scenarios (*Lebrun et al.*, 2013; *Solomatov et al.*, 1993). We further calculate the diapir sinking speed as a function of diapir size and mantle viscosity according to Stokes Law, and thereby estimate the depth, to which diapirs sink before they thermally equilibrate (Fig. 3d). Note that the scale of final-stage overturns in our geodynamic models (e.g. Fig. 4) remains an overestimate, particularly as $v_{MOC}$ is assumed to be constant (cf. Fig. 3b).

We find that any diapirs that have been formed during the final stages of MO freezing in the uppermost mantle (i.e., the top ~100 km or ~300 km, depending on MO cooling scenario) are very small (i.e. 0.1 km to 10 km, depending on MO cooling scenario). Thus they sink very slowly through the hot underlying cumulates, i.e. too slowly to avoid thermal equilibration before they reach the CMB (Fig. 9b). Under a range of realistic conditions, diapirs equilibrate before they even reach the lower mantle (i.e., their sinking depth is <660 km). These very Fe-rich diapirs have crystallized at much lower potential temperatures compared to the underlying proto-mantle due to the effects of composition on the solidus. In other words, the cumulate geotherm in our models is strongly super-adiabatic (i.e. by ~1100 K or more, depending on case). Thus, Fe-rich "proto-crustal" diapirs experience melting as soon as they thermally equilibrate. Indeed, we do not expect the strongly Fe-rich final-stage cumulates (or "proto crust") to have survived overturn, and to be directly represented in the present-day lowermost mantle.

Instead, we infer that diapirs should induce mantle hybridization and metasomatism as they undergo melting during sinking through the near-pyrolitic proto-mantle (Fig. 10). As diapirs undergo (partial) melting, the related Fe-rich melts should infiltrate the ambient pyrolitic proto-mantle. Co-existence of Fe-rich silicate melts and pyrolitic solids is thermodynamically unstable, such that hybrid lithologies with intermediate $Fe\#_{solid}$ will form. In addition, proto-crustal diapirs may have been enriched in silica, e.g. if the final-stage MO is buffered towards basaltic or komatiitic compositions due to partial melting of the convecting upper-mantle cumulates (grey dots in Fig. 10). Any such Si-enrichment of proto-crustal diapirs (*Carlson et al.*, 2015) may lead to the formation of pyroxenites during hybridization, analogous to the experimentally-supported hybridization of peridotites upon infiltration by silicic melts (*Mallik and Dasgupta*, 2012; *Yaxley and Green*, 1998).

The lithological assemblages resulting from hybridization should consist of veins of moderately Fe-enriched (and perhaps Si-enriched) hybrid lithologies in a matrix of pyrolite. Such lithological assemblages would involve density anomalies that are significantly smaller than those of the original strongly Fe-enriched cumulates. Therefore, any predicted $\Delta\rho_{comp}$ for scenarios (F), (IF) and (EIF) are to be corrected downward. In particular, such a correction will move the predictions for scenarios (IF) and (EIF) into a zone, where model predictions in terms of $\Delta\rho_{comp}$ and $V_{unmixed}$ can reconcile present-day LLSVP properties (golden shading in Fig. 6). Accordingly, our analysis of diapir sinking depths (Fig. 9b) lends credibility to a switch from intermediate to fractional crystallization during MO freezing somewhere in the upper mantle (such as in scenarios (IF) and (EIF)), consistent with inferences from MO cooling histories (see section 3). Most importantly, hybridization of the most Fe-enriched cumulates during sinking can prevent the formation of stable stratification within the deep dense layer (e.g., see blue layer in Figure 8), which is inconsistent with seismic constraints for the lowermost mantle.

Along these lines, the present-day LLSVPs may be made up of ancient/primordial lithological assemblages that involve (veins of) Fe-rich hybrid lithologies in a matrix of proto-mantle rocks. Significant enrichment of Fe within LLSVPs is consistent with seismic observations (*Mosca et al.*, 2012; *Trampert et al.*, 2004). A possible Si-enrichment of these lithological assemblages (see above) may further reconcile the anti-correlation of bulk-sound and shear-wave speed that has been localized within LLSVPs by joint body-wave seismic tomography (*Koelemeijer et al.*, 2016; *Moulik and Ekström*, 2016; *Tesoniero et al.*, 2015). *Deschamps et al.* (2012) demonstrate that a combination of moderate Fe-enrichment and Si-enrichment can indeed sustain this anti-correlation.

This analysis demonstrates that the large-scale homogeneity of the Earth's mantle (i.e., with only moderate compositional heterogeneity such as LLSVPs, and no long-lived stratification − or layered mantle convection) may ultimately be related to the presence of a thick steam atmosphere during MO freezing. Such a steam atmosphere has likely promoted slow cooling of the MO on Earth (see section 3), and sustained the small diapir sizes of final-stage cumulate overturns that are required for hybridization and/or dilution of the final-stage MO by moderately Fe-enriched basaltic-to-komatiitic partial melts. Small terrestrial planets such as Mars may instead have experienced much faster MO freezing (see *Mezger et al.*, 2013 and references therein), and therefore potentially exhibit long-lived mantle compositional layering. At least a moderately-thick stratified basal layer such as predicted by our models (Figs. 6, 8) − as well as by a similar study for Mars (*Maurice et al.*, 2017) − may be crucial to delay planetary cooling, and to explain the protracted history of Martian volcanism (cf. *Plesa et al.*, 2014).

## 6. Discussion

Our model assumptions imply that, initially, the whole mantle has been completely molten to form a deep global magma ocean. This initial condition is motivated by the giant-impact hypothesis for the generation of the Earth's moon (*Canup*, 2012; *Ćuk and Stewart*, 2012;

*Hartmann and Davis*, 1975). A giant impact is thought to have been sufficiently energetic to melt most of the mantle (*Nakajima and Stevenson*, 2015). Alternatively, the Moon may have been formed by the debris from multiple large (but not giant) impacts (*Rufu et al.*, 2017), potentially without complete melting of the Earth's mantle. This hypothesis implies a moderate initial depth extent of the MO, which is compatible with (but not demanded by (*Deguen et al.*, 2011; *Fischer et al.*, 2017)) the moderately-incompatible element budget of the Earth (*Wood et al.*, 2006). A limited extent of the last major MO episode can address the preservation of mantle $^{182}W/^{184}W$ heterogeneity through the Archean (*Brown et al.*, 2014; *Touboul et al.*, 2012) and up to the present-day (*Mundl et al.*, 2017). Note that any such "primordial" $^{182}W/^{184}W$ heterogeneity must have originally been formed earlier than ~25 Myrs after solar-system formation, i.e. well before lunar accretion (*Touboul et al.*, 2009).

Even though our model initial conditions imply complete melting of the mantle, our model predictions remain applicable for initial MO depth extents that are significantly smaller than that of the mantle. In particular, scenarios (EI) and (EIF) are also strictly applicable for cases with MO depth extents ≥1450 km. For example, scenario (EI) may represent a situation, in which (I) intermediate crystallization is dominant throughout the MO, but the initial MO depth extent is ~1450 km, i.e. similar to the estimate by *Wood et al.* (2006).

In case the initial depth extent of the MO was indeed significantly smaller than that of the CMB, the classical scenario for the formation of a "primary" basal magma ocean (BMO) is called into question. In this scenario, a BMO is separated from the "surficial" MO due to an isochemical density crossover between solids and liquids and/or a crossover in the slopes of the mantle solidus and adiabat (*Andrault et al.*, 2017; *Thomas et al.*, 2012). As any potential crossovers are inferred to occur at very high pressures, if at all (*Andrault et al.*, 2012; *de Koker et al.*, 2013; *Ghosh and Karki*, 2016; *Nomura et al.*, 2011; *Stixrude et al.*, 2009), they may only be relevant for (near-)complete melting of the early Earth's mantle, e.g. in a giant-impact scenario.

In any case, our model predictions imply that a secondary BMO may be formed at the base of the mantle after incremental overturns of Fe-enriched cumulates. Fe-rich cumulates should undergo efficient melting once they reach the base of the mantle, independent of any potential hybridization during sinking (see above). Melting in the hot thermal boundary layer near the CMB is expected in the hot Hadean eon (e.g., *Korenaga*, 2006) for any material with significant

Fe-enrichment. Note that even today, moderately-enriched rocks such as basalts are thought to almost reach melting temperatures at the CMB (*Andrault et al.*, 2014; *Kato et al.*, 2016). A secondary BMO made up of (hybridized) Fe-rich cumulates would be gravitationally stable due to its intrinsic density anomaly, independent of the postulated density crossover between solids and liquids (*Andrault et al.*, 2017). If a primary BMO with near-pyrolitic composition indeed existed before incremental overturns of MO cumulates, the strong Fe-enrichment of final-stage cumulates may be further diluted in a mixed primary-secondary BMO, i.e. in addition to the predicted hybridization during sinking (see section 5.3).

Thus, the rather short-lived manifestation of a surficial MO may implicitly lead to the formation of a long-lived secondary (or mixed primary-secondary) BMO (see also *Labrosse et al.*, 2015). The typical timescales of BMO crystallization are on the order of billion years due to thermal isolation from the surface by the solid mantle (*Labrosse et al.*, 2007). In case multiple large-scale or giant-scale impacts have occurred on the early Earth (*Rubie et al.*, 2015; *Rufu et al.*, 2017), consistent with n-body simulations of planetary accretion (*Jacobson and Morbidelli*, 2014; *Kaib and Cowan*, 2015), multiple MO episodes would have episodically fed the BMO through (incremental) cumulate overturn. In this scenario, an Fe-enriched "proto-BMO" would have been stabilized before the last MO episode, hence providing a stably-stratified reservoir for primordial $^{182}W/^{184}W$ heterogeneity − even in case complete mantle melting has occurred in the aftermath of the Moon-forming giant impact (*Nakajima and Stevenson*, 2015). Along these lines, the volume and properties of the present-day LLSVPs, which are thought to consist of BMO cumulates (*Labrosse et al.*, 2007), may bear witness of a protracted history of planetary accretion, core formation, MO fractionation, and cumulate overturn.

The nature and origin of the BMO has important geochemical implications. In case the BMO indeed originates from Fe-enriched MO cumulates that have crystallized near the surface and sunk to the base of the mantle, they may be good candidates to host an early enriched reservoir (*Boyet and Carlson*, 2005), because of partitioning of much of the bulk-silicate-Earth's incompatible-element budget into the final-stage MO cumulates (*Brown et al.*, 2014). Even though any evidence for a strongly-enriched (and "hidden") early reservoir as originally proposed by *Boyet and Carlson* (2005) has been recently challenged (*Burkhardt et al.*, 2016), at least a moderately-enriched early reservoir is indeed required by the preservation of ancient isotopic mantle heterogeneity through the Archean (*Rizo et al.*, 2016; *Touboul et al.*, 2012) and up to the present-day (*Mundl et al.*, 2017). Since the overturned MO cumulate layers have never communicated with the surface except in the presence of a dense early steam atmosphere, a "secondary" BMO and the related present-day LLSVPs may further host ancient volatiles such as primordial noble gases (*Caracausi et al.*, 2016; *Mukhopadhyay*, 2012), particularly as multiple BMO episodes (see above) are considered (*Tucker and Mukhopadhyay*, 2014). Integration of geophysical and geochemical data in a self-consistent geodynamic framework will be needed to map out primordial geochemical reservoirs in the present-day mantle.

## 7. Conclusion

Because of slow cooling in the presence of a steam atmosphere, fractional crystallization is very likely to have been the dominant mode of MO freezing, at least during the late stage of freezing in the uppermost mantle. For fractional crystallization, strong enrichment upwards occurs in the cumulate layers due to partitioning of iron during crystallization. Although incremental overturns largely remix the related major-element heterogeneity through most of the mantle, the most Fe-enriched final cumulate layers remain unperturbed, and hence may potentially persist at the base of the mantle through the age of the Earth. Our scaling analysis, however, predicts that the most Fe-enriched cumulates turn over as small diapirs, and thus thermally equilibrate at rather shallow depths to undergo melting during sinking. The resulting metasomatic rock assemblages consisting of a mixture of ambient-mantle rocks and hybrid lithologies with moderate Fe-content should have accumulated at the base of the mantle. Thus, the present-day LLSVPs may be made up of these rock assemblages, consistent with their seismic characteristics.

## Figure captions

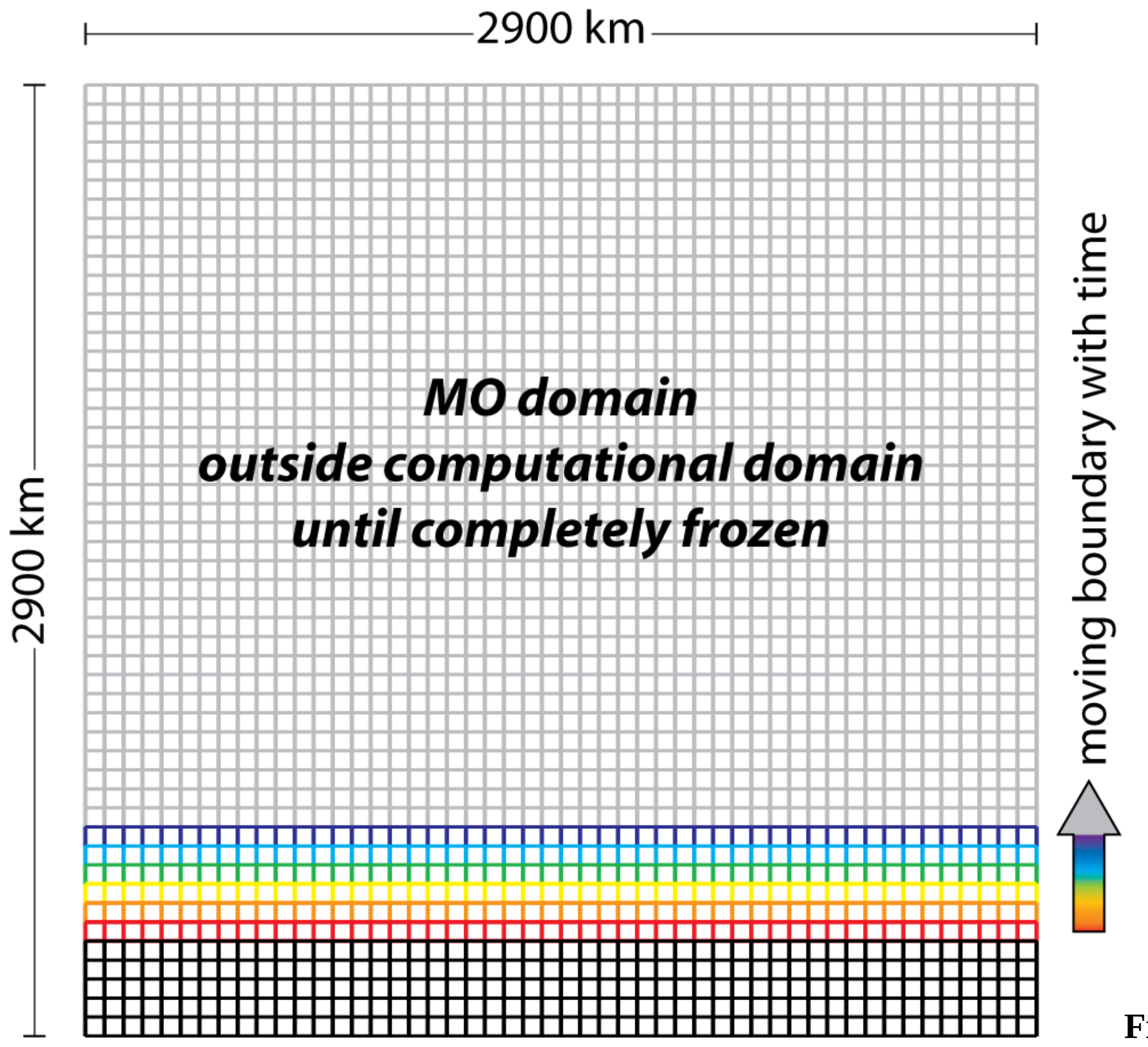

**Fig. 1:** Illustration of the numerical-model setup. The model domain of 288x256 elements is initially divided into the cumulate domain (i.e., in the fractional-crystallization scenario 26x256 elements and growing with time; schematically shown as black and colored squares) and MO domain (grey). The MO is not explicitly modeled; its thermal and compositional evolution (as a function of time, or MO depth extent $d_{MO}$) is parameterized (see Fig. 2). Thus, the MO domain acts as a top thermal and compositional boundary condition for the cumulate domain. The cumulate domain grows linearly with time as the MO freezes from below (time is schematically color-coded). The moving boundary progresses with a velocity $v_{MOC}$.

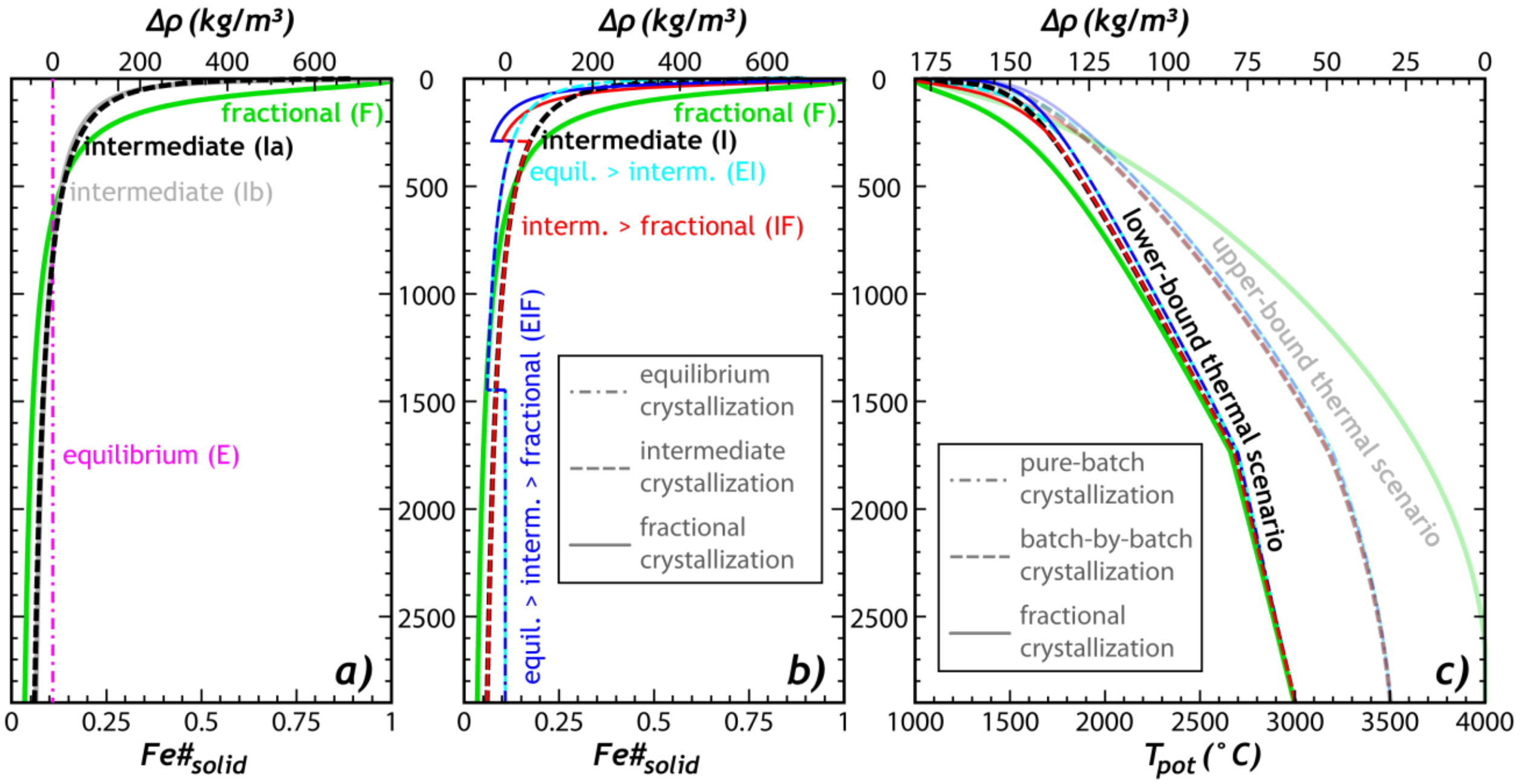

**Fig. 2:** Initial mantle (a-b) compositional profiles and (c) thermal profiles for various MO crystallization scenarios (see text). (a) Compositional profiles for end-member modes fractional crystallization and equilibrium crystallization, as well as two variants of an intermediate crystallization mode (see text). Both these intermediate variants yield very similar mantle compositional profiles. (b) Profiles for the five explicit scenarios of MO freezing explored in this study. Some scenarios consider switches between crystallization modes at 1450 and/or 290 km depth. Temperatures in (c) give MO freezing temperatures as a function of MO depth extent $d_{MO}$ for the five MO crystallization scenarios shown in (b). Opaque lines show the lower-bound (solidus) thermal scenario; transparent lines show the upper-bound thermal scenario. The kink in the solidus curve at 1740 km depth is an artifact of our bi-linear parameterization (see method section). Note that temperatures are reported as potential temperatures.

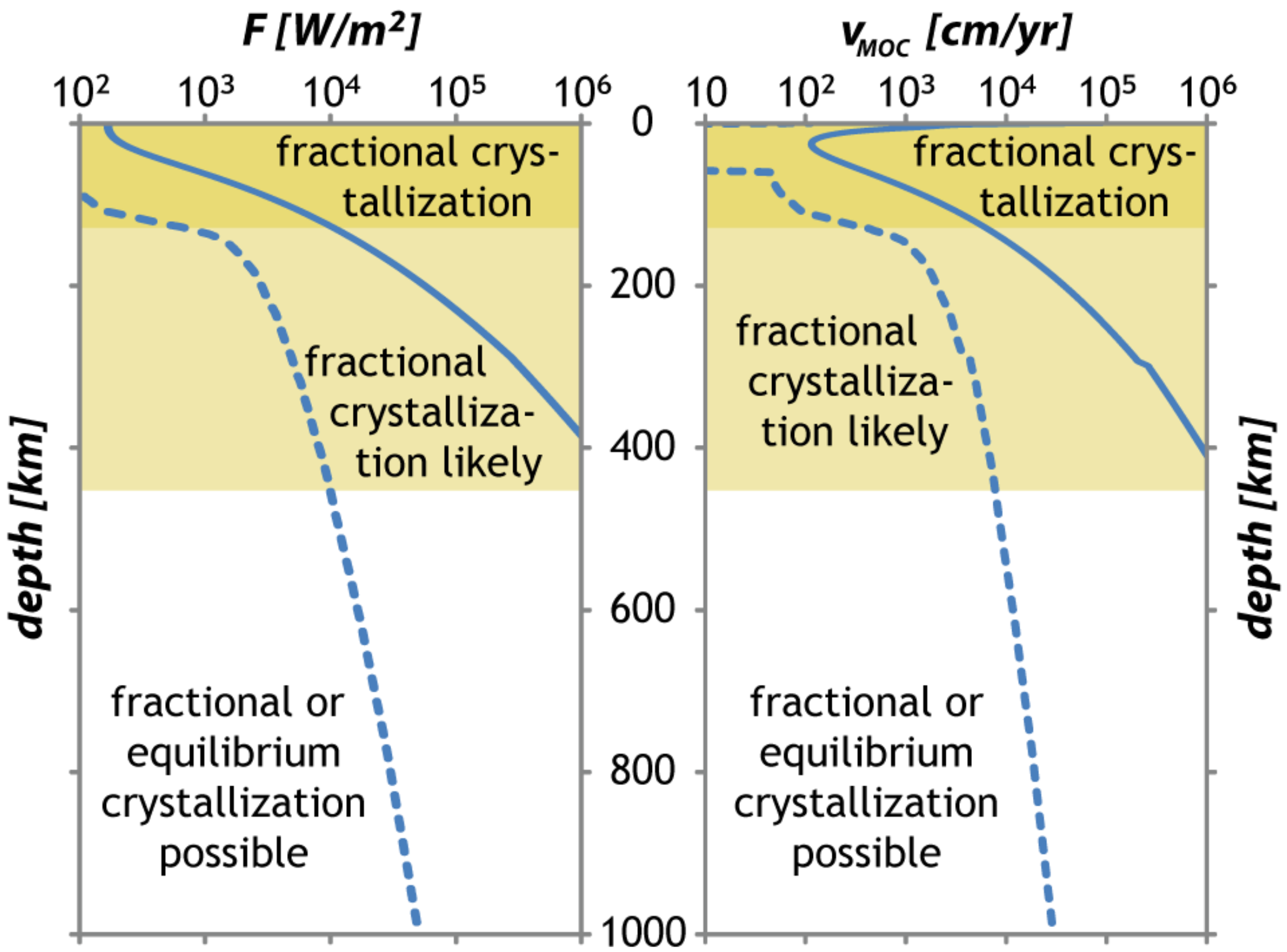

**Fig. 3**: Heat fluxes out of the MO and progression rates of the MO crystallization front. The (a) heat flux out of the MO $F$ depends on MO temperature and thus on MO depth $d_{MO}$ (we here consider the upper-bound thermal scenario $I_{290}F$). $F$ is bracketed by the dashed (*Lebrun et al.*, 2013) and solid lines (*Solomatov et al.*, 1993), systematically decreasing during MO freezing as $d_{MO}$ decreases (upward). Accordingly, the (b) predicted progression rate of the crystallization front $v_{MOC}$ also decreases (latent heat of fusion is taken to be 560 kJ/kg). Assuming that the transitional heat flux between equilibrium/intermediate and fractional crystallization $F_{crit}$ is $10^4$ W/m$^2$ < $F_{crit}$ < $10^6$ W/m$^2$ (*Solomatov*, 2015), fractional crystallization very likely dominates at least for $d_{MO}$ < 150 km, or potentially over a much greater depth range (see text).

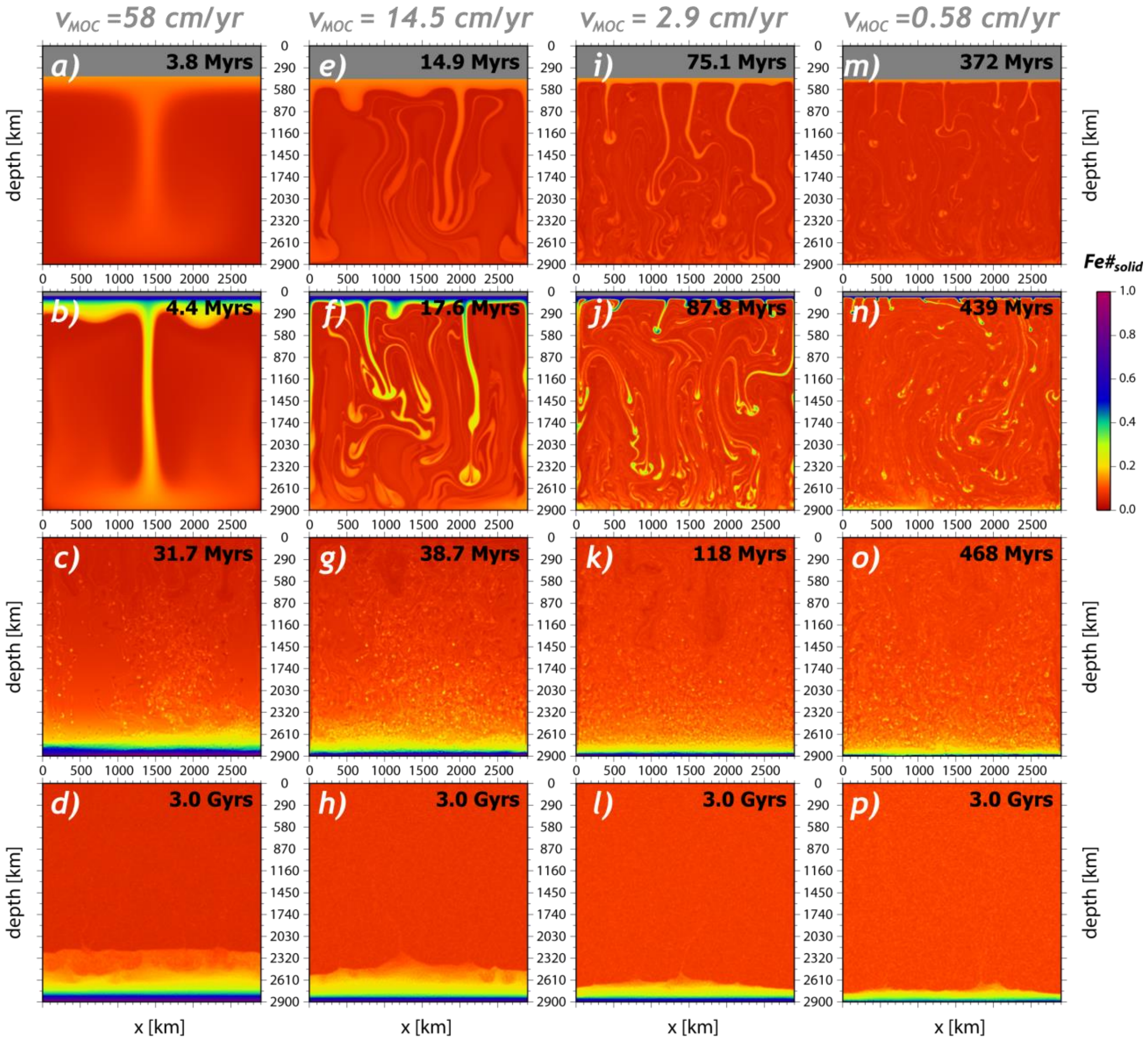

**Fig. 4**: Snapshots of MO cumulate composition for a subset of fractional-crystallization cases (scenario (F)) with variable $v_{MOC}$ (as labeled) during overturn (top two rows), just after overturn (third row), and after 3 Gyrs of solid-state mantle convection (bottom row). Grey colors mark the liquid magma-ocean domain during the late stages of crystallization in the top row. For all cases shown (1_HRa8, 1_HRa6, 1_HRa4 and 1_HRa2), $\eta_{mantle} = 1.17 \cdot 10^{20}$ Pa·s, and the lower-bound thermal scenario is applied. Smaller scales of boundary layer instability during MO crystallization for smaller $v_{MOC}$ (from left to right) cause more efficient cumulate mixing, and thus tend to decrease $V_{unmixed}$ (bottom row).

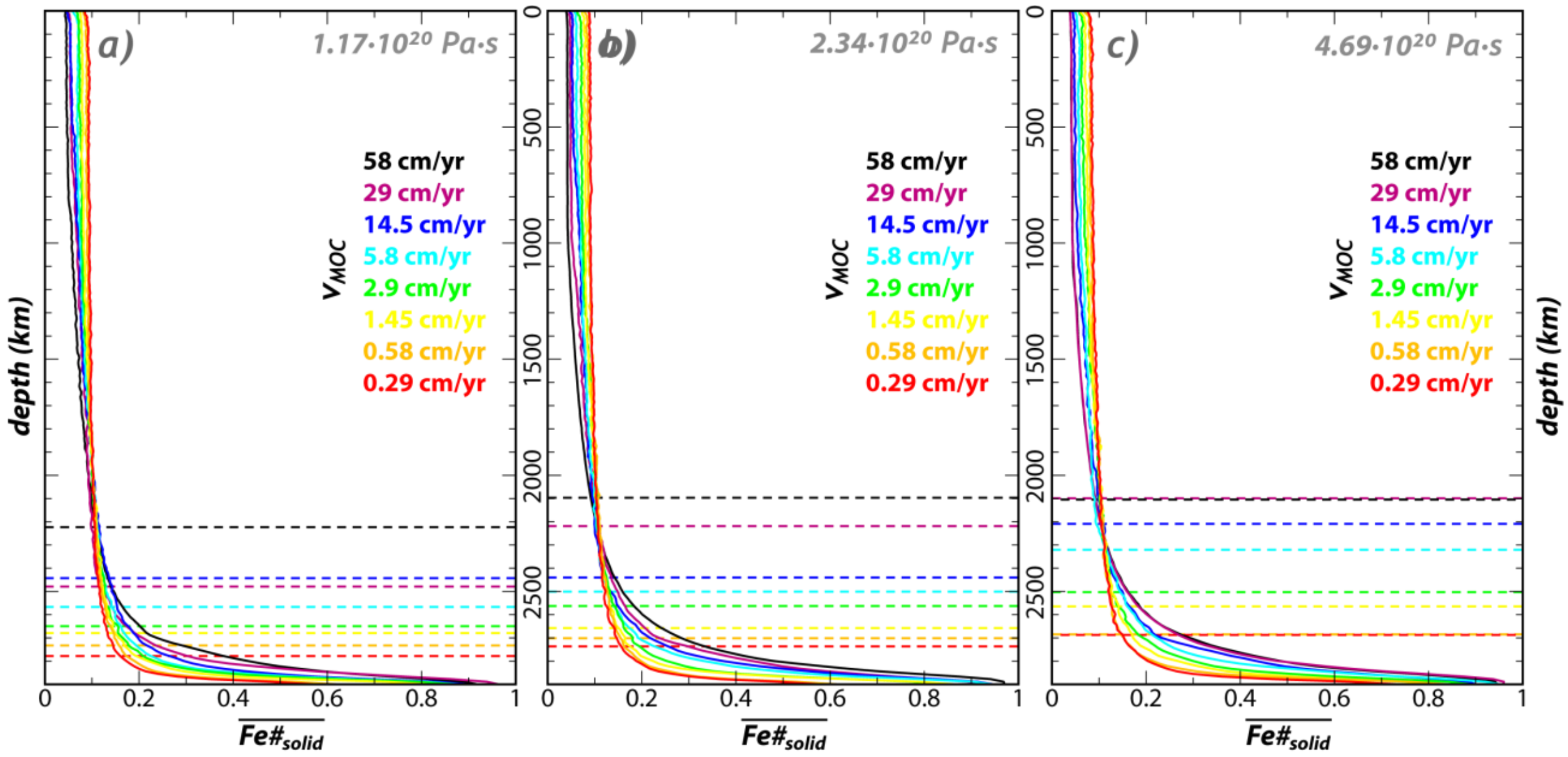

**Fig. 5:** Mantle compositional profiles just after gravitational overturn of cumulate layers for all fractional-crystallization cases in the lower-bound (solidus) thermal scenario. Curves provide compositional profiles for fractional-crystallization cases as a function of $v_{MOC}$ for (**a**) $\eta_{mantle}$ = $1.17{\cdot}10^{20}$ Pa·s (**b**) $\eta_{mantle} = 2.34{\cdot}10^{20}$ Pa·s, and (**c**) $\eta_{mantle} = 4.69{\cdot}10^{20}$ Pa·s. Dashed horizontal lines represent $V_{unmixed}$, showing the final average depth extent of the unmixed layer after 3 Gyrs.

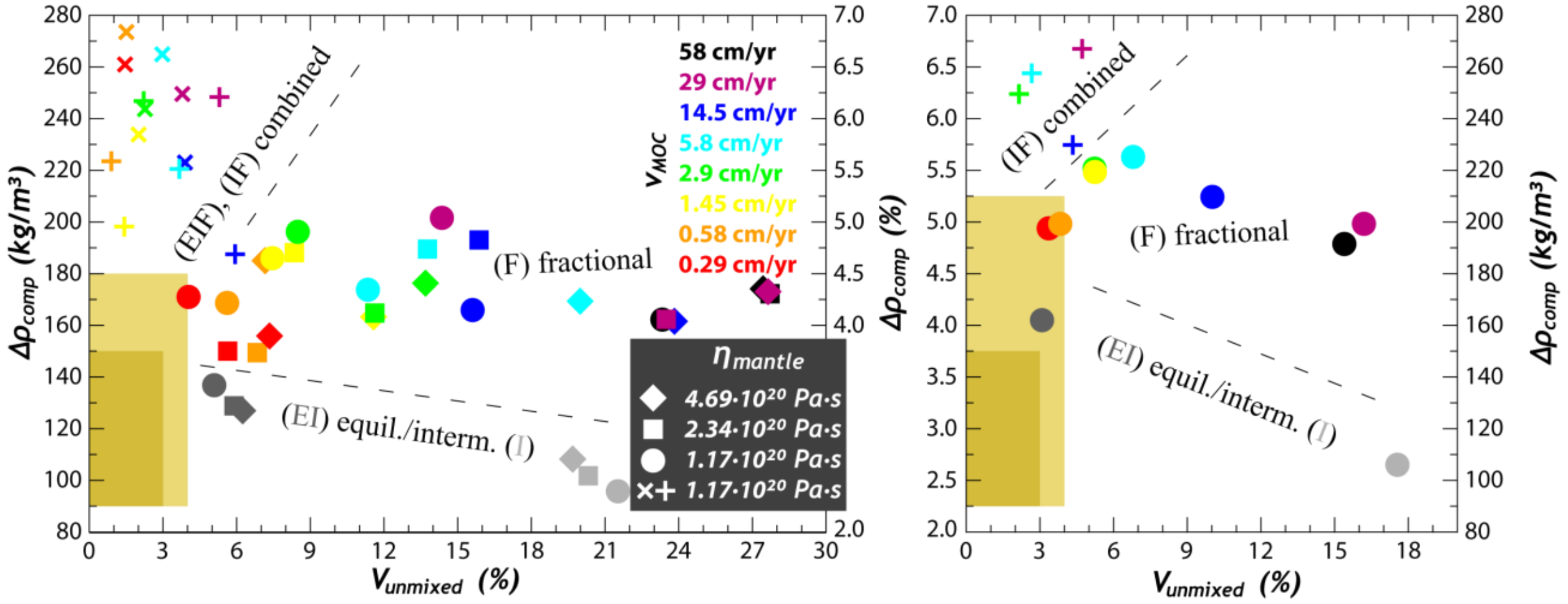

**Fig. 6**: Volume $V_{unmixed}$ and density anomaly $\Delta\rho_{comp}$ of the unmixed layer at the base of the mantle after 3 Gyrs model time for all cases for the (left) lower-bound and (right) upper-bound thermal scenario. Colored symbols (diamonds, squares, circles) are fractional-crystallization cases with $v_{MOC}$ and $\eta_{mantle}$ as labeled. $v_{MOC}$ and $\eta_{mantle}$ control the number of incremental overturns, and hence the extent of mixing. Grey symbols are equilibrium and/or intermediate crystallization cases (light grey: scenario (I), dark grey: scenario (EI)). Colored crosses are combined crystallization cases (+: scenario 3a, ×: scenario 3c). The golden rectangle shows the acceptable range of $V_{unmixed}$ and $\Delta\rho_{comp}$ as the constraint from LLSVP volumes and shapes is taken into account. The light-golden shading symbolizes that somewhat softer bounds may apply as various additional processes are taken into account (see text).

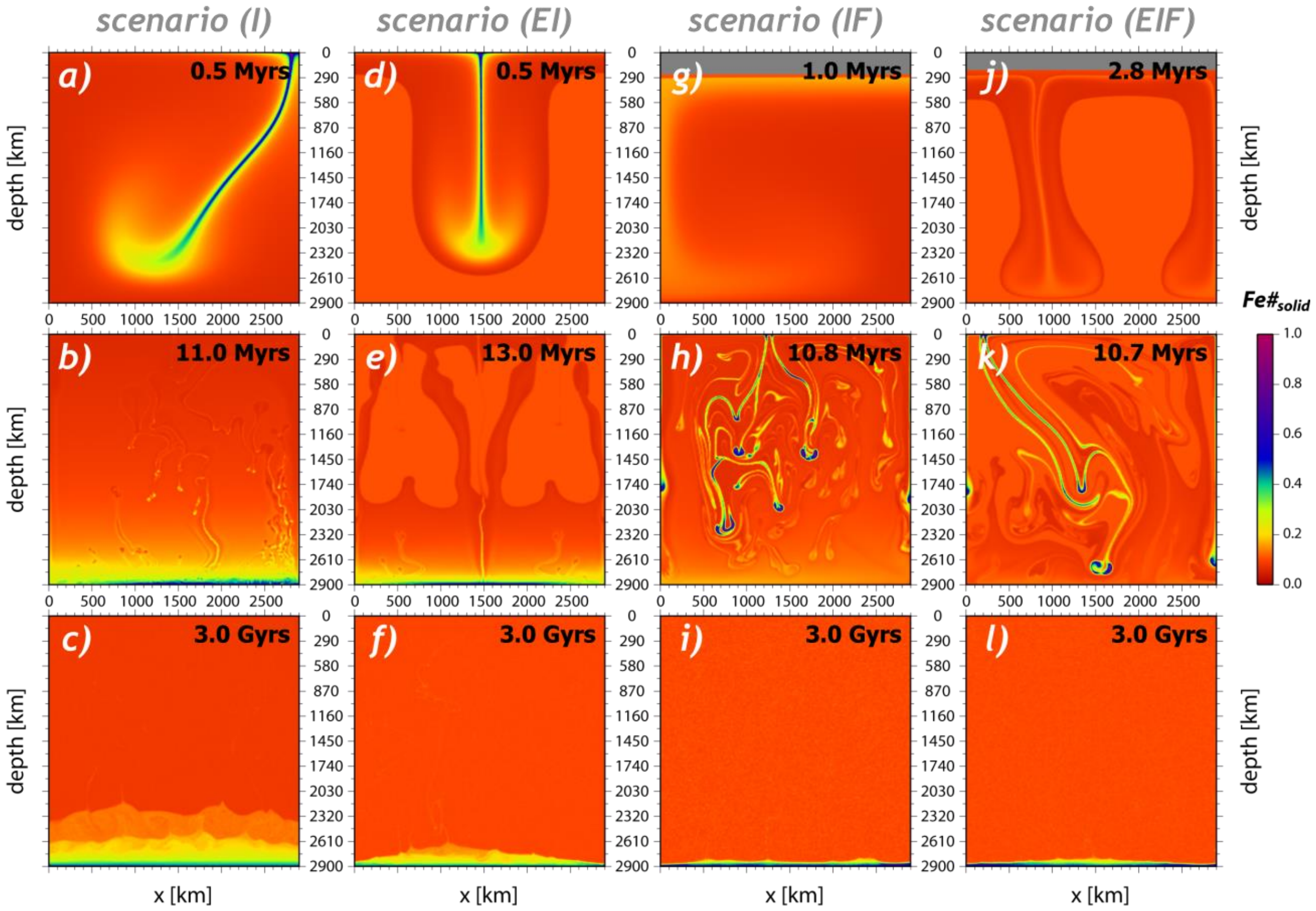

**Fig. 7**: Snapshots of mantle composition with model times as annotated for intermediate ((I), (EI)) and combined ((IF), (EIF)) crystallization scenarios. Initial compositional profiles are given in Figure 2a. For all cases shown ((a) I_HRa, (b) EI_HRa, (c) IF_HRa4, and (d) EIF_HRa4), $\eta_{mantle} = 1.17 \cdot 10^{20}$ Pa·s, and the lower-bound thermal scenario is applied. In combined-crystallization cases IF_HRa4 and EIF_HRa4, $v_{MOC} = 2.9$ cm/yr. The corresponding fractional-crystallization case with $\eta_{mantle} = 1.17 \cdot 10^{20}$ Pa·s and $v_{MOC} = 2.9$ cm/yr is shown in Figures 4i-l.

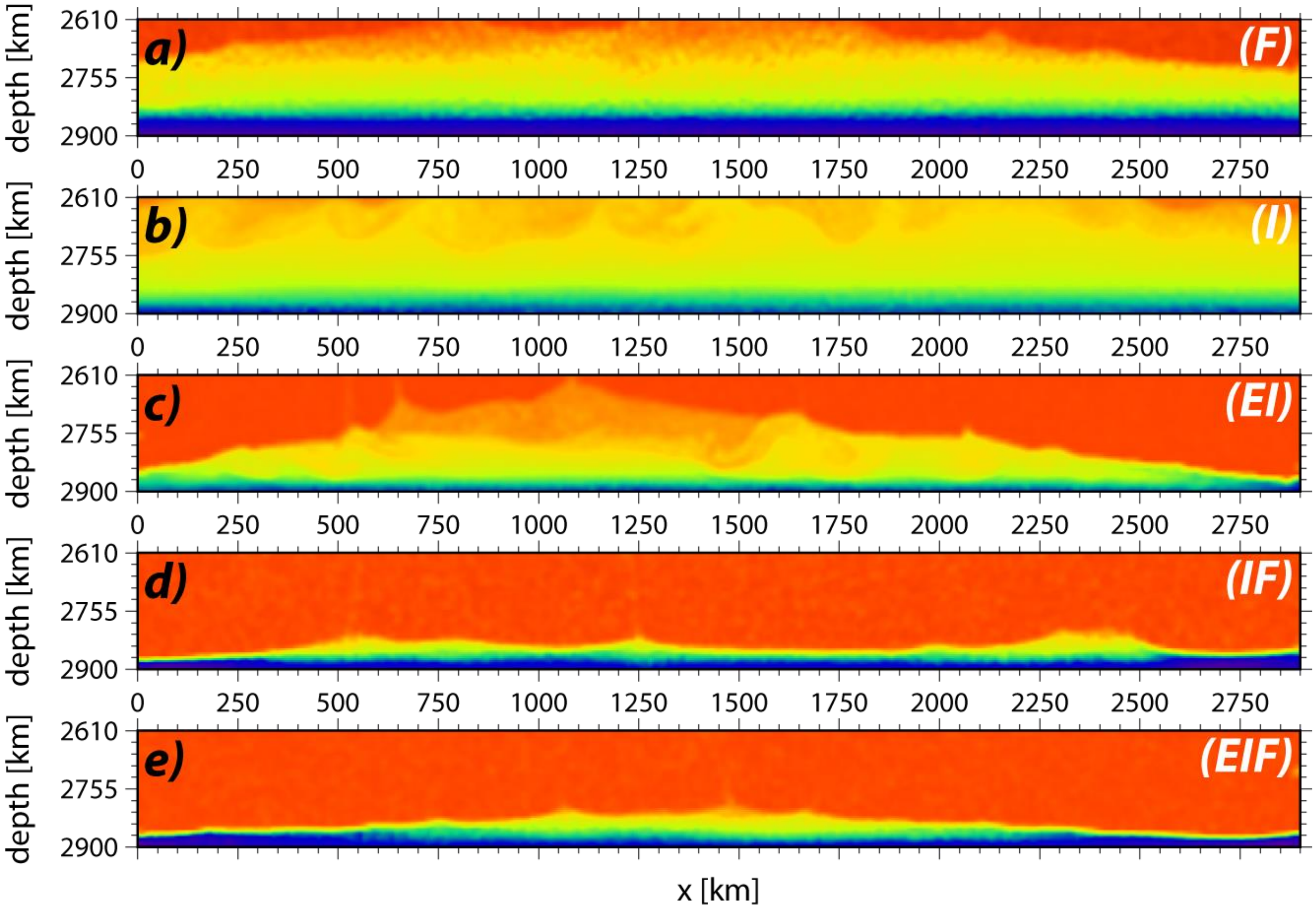

**Fig. 8**: Compositional stratification in the lowermost mantle after 3 Gyrs for a subset of cases that span all MO crystallization scenarios (as annotated). Cases shown are the same as in Figures 4i-l and 7 ((a) F_HRa4, (b) I_HRa, (c) EI_HRa, (d) IF_HRa4, and (e) EIF_HRa4). Thus, the five panels ((a)-(e)) are zoom-ins into Figures 4l, 7c, 7f, 7i and 7l, respectively. For all cases, a (variably thick) strongly Fe-enriched layer is preserved at the very base of the mantle.

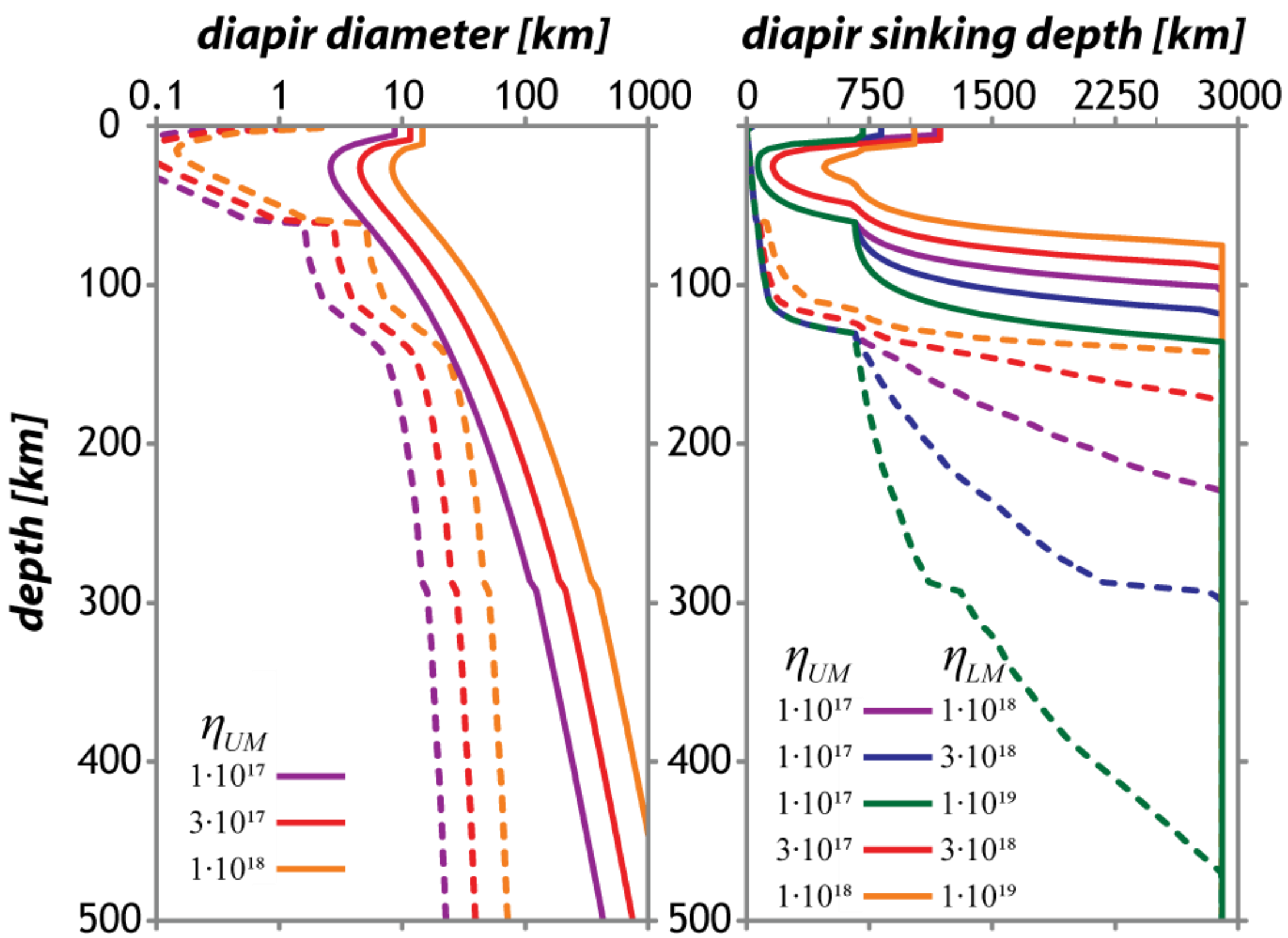

**Fig. 9**: Cumulate diapir diameters and diapir sinking depths as a function of MO depth $d_{MO}$ during the late stages of MO freezing. Diapir diameters (a) are calculated as critical thermal-boundary-layer thicknesses from $v_{MOC}$ and upper-mantle viscosity $\eta_{UM}$ (see Appendix A1-A2) and plotted versus $d_{MO}$. Note that $v_{MOC}$ is a function of $d_{MO}$ (Figure 3) for a given cooling scenario of the MO. Diapir diameters decrease from ~100 km to ~1 km through the upper mantle, bracketed by cooling scenarios (dashed line (*Lebrun et al.*, 2013); solid line (*Solomatov et al.*, 1993)). Diapir sinking depths (b), here defined as the depth of thermal equilibration during sinking, depend on diapir diameter and mantle rheology according to Stokes Law (see Appendix A2). For reasonable upper-mantle $\eta_{UM}$ and lower-mantle viscosities $\eta_{LM}$ in the early Earth, the final and most strongly Fe-enriched cumulates that are formed in the upper 100~300 km should thermally equilibrate before they reach the base of the mantle.

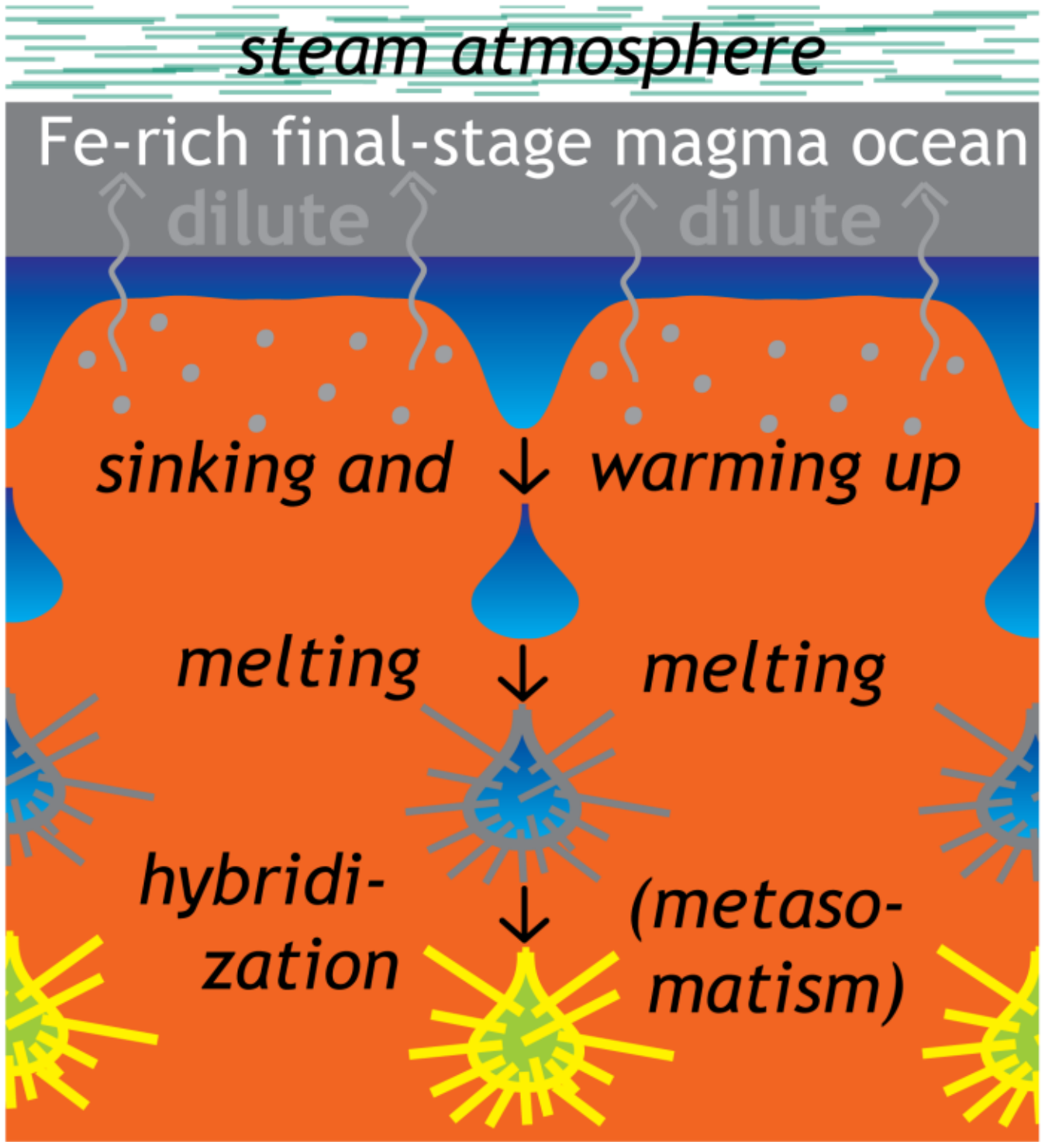

**Fig. 10**: Cartoon illustrating the fate of final-stage MO cumulates. Fe-rich cumulates that form in the shallow final-stage MO are predicted to sink as small diapirs (Fig. 9a) to thermally equilibrate (Fig. 9b) and melt during sinking. The resulting moderately Fe-enriched lithological assemblages that consist of a mix of refrozen melts, hybrid rock and ambient mantle may make up the present-day LLSVPs (see text), in particular if the final-stage MO had been modified (to balance $Fe\#_{MO}$ and to boost $SiO_2$ contents) by partial melts of the convecting cumulate layers (light grey arrows).

## Appendix A1: Rayleigh-Taylor instability for a growing boundary layer

The timescale for the growth of a Rayleigh-Taylor instability is given by (*Turcotte and Schubert*, 1982) as:

$$\tau_{RT} = \frac{13.04\eta}{\Delta\rho g b_{crit}} \quad \text{(A1)}$$

with $b_{crit}$ the critical thickness of the unstable layer, and $\Delta\rho$ the relevant density difference. In the case of a boundary layer that is growing from the backside due to the addition of material (as in the case of a freezing MO), as well as rejuvenated on the frontside by convective removal on the timescale $\tau_{RT}$, the critical boundary layer thickness becomes:

$$b_{crit} = \tau_{RT} v_{MOC} \quad \text{(A2)}$$

where $v_{MOC}$ is the progression velocity of the "backside" due to addition of material. Combining equations (A1) and (A2) gives:

$${b_{crit}}^2 = \frac{13.04\eta v_{MOC}}{\Delta\rho g} \quad \text{(A3)}$$

This relation implies that there is a direct trade-off between $\eta$ and $v_{MOC}$ in terms of critical boundary layer thicknesses for Rayleigh-Taylor instability. Increasing $\eta$ and decreasing $v_{MOC}$ both by a factor of two, or vice-versa, gives the same $b_{crit}$. Along these lines, we also establish a direct trade-off between $\eta$ and $v_{MOC}$ in terms of wavelengths of instability, which are generally proportional to $b$ (*Turcotte and Schubert*, 1982), and related mixing. Figure A1 shows that these direct trade-offs are confirmed by our numerical experiments.

The above analysis is appropriate for the instability of a purely compositional boundary layer, but it is expected hold for a thermochemical boundary layer as long as $\tau_{RT}$ is smaller or of the same order than the timescale of thermal diffusion, which is generally the case for realistic timescales of MO freezing. Also note that compositional instability (due to negative buoyancy) is much stronger than thermal instability, at least during crystallization of the shallow MO (Fig. 2), a stage which is critical for cumulate mixing.

## Appendix A2: Thermal equilibration during cumulate sinking

Any downwelling diapirs related to incremental overturns of the last and most Fe-enriched cumulate layers may thermally equilibrate during sinking through the mantle. The mantle, through which the last cumulate diapirs sink, is much warmer than the diapirs themselves, because it has crystallized at higher solidus temperatures (see Fig. 2c). The timescales of thermal equilibration $\tau_{equil}$ and those of sinking $\tau_{sink}$ depend on the size of the sinking diapirs. According to Figure 9a, diapir diameters are <10 km during the final stages of magma-ocean crystallization. These diameters are calculated from $b_{crit}$ (equation A3) assuming near-solidus upper-mantle viscosities of $10^{17}$ Pa·s < $\eta_{UM}$ < $10^{18}$ Pa·s, as well as taking thermal plus chemical density anomalies from Figure 2 (scenario $I_{290}F$) and $v_{MOC}$ from two different magma-ocean cooling models (dashed line versus solid line in Figure 9). Magma-ocean cooling models (*Lebrun et al.*, 2013; *Solomatov et al.*, 1993) are visualized in Figure 3. Accordingly, thermal equilibration occurs on timescales of about $\tau_{equil}$ < 1.5 Myrs ($\tau_{equil} = 0.25 b_{crit}^2/\kappa$). Note that thermal equilibration in our numerical models takes much longer than predicted by these estimates, as $b_{crit}$ are systematically overestimated (we apply $\eta_{mantle} > 10^{20}$ Pa·s due to computational limitations).

The timescale for sinking diapiric downwellings from the moving boundary layer through the upper and lower mantle is given by Stoke's Law as:

$$\tau_{sink} = \frac{660km \cdot 18\eta_{UM}}{\Delta\rho g {b_{crit}}^2} + \frac{2240km \cdot 18\eta_{LM}}{\Delta\rho g {b_{crit}}^2} \quad \text{(A4)}$$

As long as $\eta_{UM} < 7 \cdot 10^{18}$ Pa·s and $\eta_{LM}/\eta_{UM} \geq 10$, which is likely, particularly because the initial lower mantle is mostly composed of stiff $MgSiO_3$ bridgmanite (see main text), $\tau_{sink}$ is larger than $\tau_{equil}$. For $\eta_{UM} \leq 2 \cdot 10^{17}$ Pa·s, thermal equilibration will be even complete before diapirs reach the base of the upper mantle at 660 km depth. These bounds for early-Earth mantle viscosity are conservative estimates using the *Solomatov et al.* (1993) MO cooling model, which ignores atmospheric insulation of the MO, and would be higher using the more realistic *LeBrun et al.* (2013) MO cooling model (dashed line in Figure 9b). In any case, the last cumulates formed in the uppermost mantle (up to 100~300 km depth) should thermally equilibrate during sinking, i.e. well before they reach the lowermost mantle.

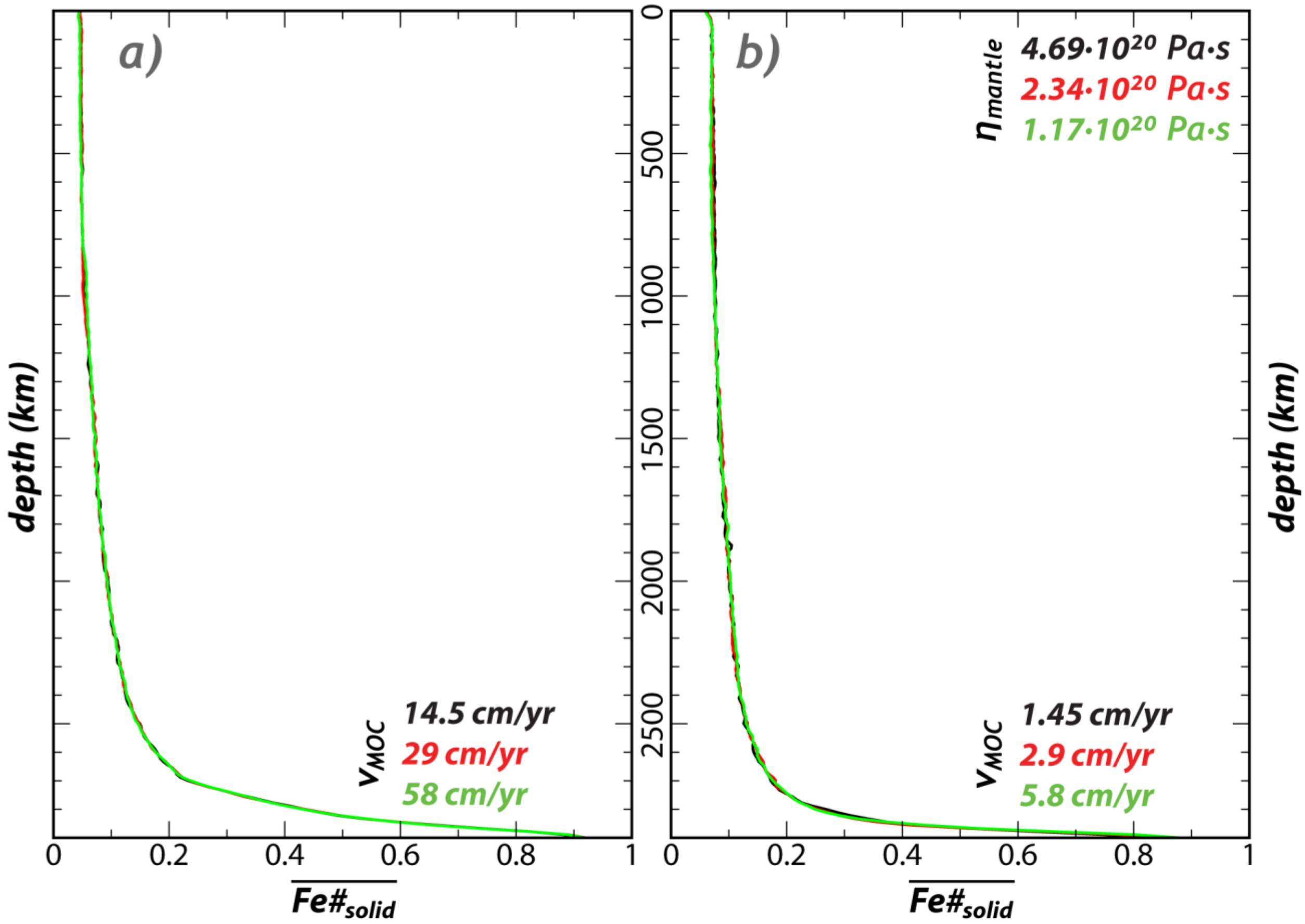

**Fig. A1:** Mantle compositional profiles soon after overturn for selected fractional-crystallization cases (as labeled). In both panels, $\eta_{mantle}$ = 4.69·10$^{20}$ Pa·s (black line); $\eta_{mantle}$ = 2.34·10$^{20}$ Pa·s (red line); $\eta_{mantle}$ = 1.17·10$^{20}$ Pa·s (green line). The similarity of all three curves (each with the same $\tau_{MOC}$ $\eta_{mantle}$) in both panels experimentally confirms the direct trade-off between $\tau_{MOC}$ and $\eta_{mantle}$ in terms of convective instability of a boundary layer that grows due to addition of material at the backside.

## Acknowledgements

We are grateful to the editor and two anonymous reviewers for comments that helped to improve the manuscript. We also thank S. Solomatov, J. Hernlund, S. Dude and M. Hirschmann for their feedback during earlier stages of the study. Panels in Figures 4, 7 and 8 have been visualized using the Generic Mapping Tools (http://gmt.soest.hawaii.edu). Calculations have been performed on the computer cluster *akua* (Univ. Hawaii). The version of CitcomCU that has been used to perform the calculations, as well as the relevant parameter files, are available at http://jupiter.ethz.ch/~ballmerm/Ballmer_et_al--GGG2017--repository.zip.

# References

Abe, Y., and T. Matsui (1986), Early evolution of the Earth: accretion, atmosphere formation, and thermal history, paper presented at Lunar and Planetary Science Conference Proceedings.

Abe, Y. (1995), Basic equations for evolution of partially molten mantle and core, in *The Earth's Central Part: Its Structure and Dynamics*, edited by T. Yukutake, pp. 215-235, TERRAPUB, Tokyo.

Abe, Y. (1997), Thermal and chemical evolution of the terrestrial magma ocean, *Phys. Earth Planet. Int.*, *100*, 27-39.

Andrault, D., N. Bolfan-Casanova, G. L. Nigro, M. A. Bouhifd, G. Garbarino, and M. Mezouar (2011), Solidus and liquidus profiles of chondritic mantle: Implication for melting of the Earth across its history, *Earth Planet. Sci. Lett.*, *304*(1), 251-259.

Andrault, D., S. Petitgirard, G. L. Nigro, J.-L. Devidal, G. Veronesi, G. Garbarino, and M. Mezouar (2012), Solid-liquid iron partitioning in Earth/'s deep mantle, *Nature*, *487*(7407), 354-357.

Andrault, D., G. Pesce, M. A. Bouhifd, N. Bolfan-Casanova, J.-M. Hénot, and M. Mezouar (2014), Melting of subducted basalt at the core-mantle boundary, *Science*, *344*(6186), 892-895.

Andrault, D., J. Monteux, M. Le Bars, and H. Samuel (2016), The deep Earth may not be cooling down, *Earth Planet. Sci. Lett.*, *443*, 195-203.

Andrault, D., N. Bolfan-Casanova, M. A. Bouhifd, A. Boujibar, G. Garbarino, G. Manthilake, M. Mezouar, J. Monteux, P. Parisiades, and G. Pesce (2017), Toward a coherent model for the melting behavior of the deep Earth's mantle, *Phys. Earth Planet. Inter.*, *265*, 67-81.

Arzi, A. A. (1978), Critical phenomena in the rheology of partially melted rocks, *Tectonophysics*, *44*(1), 173-184.

Austermann, J., B. T. Kaye, J. X. Mitrovica, and P. Huybers (2014), A statistical analysis of the correlation between large igneous provinces and lower mantle seismic structure, *Geophys. J. Int.*, *197*(1), 1-9, doi:10.1093/gji/ggt500.

Ballmer, M. D., N. C. Schmerr, T. Nakagawa, and J. Ritsema (2015), Compositional mantle layering revealed by slab stagnation at ~1000-km depth, *Science Advances*, *1*(11), doi:10.1126/sciadv.1500815.

Ballmer, M. D., L. Schumacher, V. Lekic, C. Thomas, and G. Ito (2016), Compositional layering within the large low shear-wave velocity provinces in the lower mantle, *Geochemistry, Geophysics, Geosystems*, doi:10.1002/2016GC006605.

Ballmer, M. D., C. Houser, J. W. Hernlund, R. M. Wentzcovitch, and K. Hirose (2017), Persistence of strong silica-enriched domains in the Earth's lower mantle, *Nature Geoscience*, *10*(3), 236-240, doi:10.1038/ngeo2898.

Bian, H., N. Sun, and Z. Mao (2016), Effects of chemical, mineralogical, and temperature variations on the density and bulk sound velocity of the bottom lower mantle, *Science China Earth Sciences*, *59*(10), 2062-2077.

Boukare, C. E., Y. Ricard, and G. Fiquet (2015), Thermodynamics of the MgO-FeO-SiO2 system up to 140GPa: Application to the crystallization of Earth's magma ocean, *Journal of Geophysical Research-Solid Earth*, *120*(9), 6085-6101, doi:10.1002/2015jb011929.

Boyet, M., and R. W. Carlson (2005), 142Nd evidence for early (>4.53 Ga) global differentiation of the silicate Earth, *Science*, *309*, 576-581.

Brandenburg, J. P., E. H. Hauri, P. E. van Keken, and C. J. Ballentine (2008), A multiple-system study of the geochemical evolution of the mantle with force-balanced plates and thermochemical effects, *Earth Planet. Sci. Lett.*, *276*(1-2), 1-13, doi:10.1016/j.epsl.2008.08.027.

Brown, S. M., L. T. Elkins-Tanton, and R. J. Walker (2014), Effects of magma ocean crystallization and overturn on the development of Nd-142 and W-182 isotopic heterogeneities in the primordial mantle, *Earth Planet. Sci. Lett.*, *408*, 319-330, doi:10.1016/j.epsl.2014.10.025.

Burkhardt, C., L. E. Borg, G. A. Brennecka, Q. R. Shollenberger, N. Dauphas, and T. Kleine (2016), A nucleosynthetic origin for the Earth's anomalous 142Nd composition, *Nature*, *537*(7620), 394-398, doi:10.1038/nature18956.

Canup, R. M. (2012), Forming a Moon with an Earth-like composition via a giant impact, *Science*, *338*(6110), 1052-1055.

Caracausi, A., G. Avice, P. G. Burnard, E. Füri, and B. Marty (2016), Chondritic xenon in the Earth's mantle, *Nature*, *533*(7601), 82-85, doi:10.1038/nature17434.

Carlson, R. W., M. Boyet, J. O'Neil, H. Rizo, and R. J. Walker (2015), Early Differentiation and Its Long-Term Consequences for Earth Evolution, *The Early Earth: Accretion and Differentiation*, 143-172.

Corgne, A., C. Liebske, B. J. Wood, D. C. Rubie, and D. J. Frost (2005), Silicate perovskite-melt partitioning of trace elements and geochemical signature of a deep perovskitic reservoir, *Geochimica Et Cosmochimica Acta*, *69*(2), 485-496, doi:10.1016/j.gca.2004.06.041.

Costa, A., L. Caricchi, and N. Bagdassarov (2009), A model for the rheology of particle-bearing suspensions and partially molten rocks, *Geochemistry, Geophysics, Geosystems*, *10*(3).

Cottaar, S., and V. Lekic (2016), Morphology of seismically slow lower-mantle structures, *Geophys. J. Int.*, *207*(2), 1122-1136.

Ćuk, M., and S. T. Stewart (2012), Making the Moon from a fast-spinning Earth: a giant impact followed by resonant despinning, *Science*, *338*(6110), 1047-1052.

Davaille, A. (1999), Simultaneous generation of hotspots and superswells by convection in a heterogeneous planetary mantle, *Nature*, *402*(6763), 756-760.

Davies, D. R., S. Goes, J. H. Davies, B. S. A. Schuberth, H. P. Bunge, and J. Ritsema (2012), Reconciling dynamic and seismic models of Earth's lower mantle: The dominant role of thermal heterogeneity, *Earth Planet. Sci. Lett.*, *353*, 253-269, doi:10.1016/j.epsl.2012.08.016.

Davies, D. R., S. Goes, and M. Sambridge (2015), On the relationship between volcanic hotspot locations, the reconstructed eruption sites of large igneous provinces and deep mantle seismic structure, *Earth Planet. Sci. Lett.*, *411*, 121-130, doi:10.1016/j.epsl.2014.11.052.

de Koker, N., B. B. Karki, and L. Stixrude (2013), Thermodynamics of the MgO–SiO 2 liquid system in Earth's lowermost mantle from first principles, *Earth Planet. Sci. Lett.*, *361*, 58-63.

Deguen, R., P. Olson, and P. Cardin (2011), Experiments on turbulent metal-silicate mixing in a magma ocean, *Earth*

*Planet. Sci. Lett.*, *310*(3), 303-313, doi:10.1016/j.epsl.2011.08.041.
Deschamps, F., and P. J. Tackley (2009), Searching for models of thermo-chemical convection that explain probabilistic tomography. II-Influence of physical and compositional parameters, *Phys. Earth Planet. Inter.*, *176*(1-2), 1-18, doi:10.1016/j.pepi.2009.03.012.
Deschamps, F., L. Cobden, and P. J. Tackley (2012), The primitive nature of large low shear-wave velocity provinces, *Earth Planet. Sci. Lett.*, *349–350*(0), 198-208, doi:10.1016/j.epsl.2012.07.012.
Dobson, D. P., and J. P. Brodholt (2005), Subducted banded iron formations as a source of ultralow-velocity zones at the core–mantle boundary, *Nature*, *434*(7031), 371-374.
Du, Z., and K. K. M. Lee (2014), High-pressure melting of MgO from (Mg,Fe)O solid solutions, *Geophys. Res. Lett.*, *41*(22), 8061-8066, doi:10.1002/2014GL061954.
Elkins-Tanton, L. T., P. C. Hess, and E. M. Parmentier (2005), Possible formation of ancient crust on Mars through magma ocean processes, *J. Geophys. Res.*, *110*(E12S01), doi:10.1029/2005JE002480.
Elkins-Tanton, L. T. (2008), Linked magma ocean solidification and atmospheric growth for Earth and Mars, *Earth Planet. Sci. Lett.*, *271*(1-4), 181-191, doi:10.1016/j.epsl.2008.03.062.
Fiquet, G., A. L. Auzende, J. Siebert, A. Corgne, H. Bureau, H. Ozawa, and G. Garbarino (2010), Melting of Peridotite to 140 Gigapascals, *Science*, *329*(5998), 1516-1518, doi:10.1126/science.1192448.
Fischer, R. A., A. J. Campbell, and F. J. Ciesla (2017), Sensitivities of Earth's core and mantle compositions to accretion and differentiation processes, *Earth Planet. Sci. Lett.*, *458*, 252-262, doi:10.1016/j.epsl.2016.10.025.
Flasar, F. M., and F. Birch (1973), Energetics of core formation: a correction, *J. geophys. Res*, *78*, 6101-6103.
Forsyth, D. W., D. S. Scheirer, S. C. Webb, et al. (1998), Imaging the deep seismic structure beneath a mid-ocean ridge: The MELT experiment, *Science*, *280*(5367), 1215-1218.
Garnero, E. J., J. Revenaugh, Q. Williams, T. Lay, and L. H. Kellogg (1998), Ultralow velocity zone at the core-mantle boundary, *The core-mantle boundary region*, 319-334.
Garnero, E. J., and A. K. McNamara (2008), Structure and dynamics of Earth's lower mantle, *Science*, *320*(5876), 626-628, doi:10.1126/science.1148028.
Genda, H., and Y. Abe (2003), Survival of a proto-atmosphere through the stage of giant impacts: the mechanical aspects, *Icarus (USA)*, *164*(1), 149-162.
Ghosh, D. B., and B. B. Karki (2016), Solid-liquid density and spin crossovers in (Mg, Fe)O system at deep mantle conditions, *Scientific Reports*, *6*, 37269, doi:10.1038/srep37269.
Grand, S. P., R. D. van der Hilst, and S. Widiyantoro (1997), High resolution global tomography: a snapshot of convection in the Earth, *Geological Society of America Today*, *7*(4).
Hamano, K., Y. Abe, and H. Genda (2013), Emergence of two types of terrestrial planet on solidification of magma ocean, *Nature*, *497*(7451), 607-610.
Harrison, T. M. (2009), The Hadean crust: evidence from> 4 Ga zircons, *Annual Review of Earth and Planetary Sciences*, *37*, 479-505.
Hartmann, W. K., and D. R. Davis (1975), Satellite-sized planetesimals and lunar origin, *Icarus (USA)*, *24*(4), 504-515.
Hernlund, J. W., and C. Houser (2008), The statistical distribution of seismic velocities in Earth's deep mantle, *Earth Planet. Sci. Lett.*, *265*(3-4), 423-437, doi:10.1016/j.epsl.2007.10.042.
Hernlund, J. W., and A. K. McNamara (2015), 7.11 - The Core–Mantle Boundary Region A2 - Schubert, Gerald, in *Treatise on Geophysics (Second Edition)*, edited, pp. 461-519, Elsevier, Oxford, doi:10.1016/B978-0-444-53802-4.00136-6.
Jacobson, S. A., and A. Morbidelli (2014), Lunar and terrestrial planet formation in the Grand Tack scenario, *Phil. Trans. R. Soc. A*, *372*(2024), 20130174.
Kaib, N. A., and N. B. Cowan (2015), The feeding zones of terrestrial planets and insights into Moon formation, *Icarus (USA)*, *252*, 161-174, doi:10.1016/j.icarus.2015.01.013.
Kato, C., K. Hirose, R. Nomura, M. D. Ballmer, A. Miyake, and Y. Ohishi (2016), Melting in the FeO SiO 2 system to deep lower-mantle pressures: Implications for subducted Banded Iron Formations, *Earth Planet. Sci. Lett.*, *440*, 56-61.
Koelemeijer, P., J. Ritsema, A. Deuss, and H. J. van Heijst (2016), SP12RTS: a degree-12 model of shear- and compressional-wave velocity for Earth's mantle, *Geophys. J. Int.*, *204*(2), 1024-1039, doi:10.1093/gji/ggv481.
Korenaga, J. (2006), Archean Geodynamics and the Thermal Evolution of Earth, in *Archean Geodynamics and Environments*, edited, pp. 7-32, American Geophysical Union, doi:10.1029/164GM03.
Labrosse, S., J. W. Hernlund, and N. Coltice (2007), A crystallizing dense magma ocean at the base of the Earth's mantle, *Nature*, *450*, 866-869, doi:10.1038/nature06355.
Labrosse, S., J. W. Hernlund, and K. Hirose (2015), Fractional Melting and Freezing in the Deep Mantle and Implications for the Formation of a Basal Magma Ocean, *The Early Earth: Accretion and Differentiation*, 123-142.
Lebrun, T., H. Massol, E. Chassefière, A. Davaille, E. Marcq, P. Sarda, F. Leblanc, and G. Brandeis (2013), Thermal evolution of an early magma ocean in interaction with the atmosphere, *Journal of Geophysical Research: Planets*, *118*(6), 1155-1176, doi:10.1002/jgre.20068.
Li, M., A. K. McNamara, and E. J. Garnero (2014a), Chemical complexity of hotspots caused by cycling oceanic crust through mantle reservoirs, *Nature Geosci*, *7*(5), 366-370, doi:10.1038/ngeo2120.
Li, Y., F. Deschamps, and P. J. Tackley (2014b), The stability and structure of primordial reservoirs in the lower mantle: insights from models of thermochemical convection in three-dimensional spherical geometry, *Geophys. J. Int.*, *199*(2), 914-930.
Liebske, C., A. Corgne, D. J. Frost, D. C. Rubie, and B. J. Wood (2005), Compositional effects on element partitioning between Mg-silicate perovskite and silicate melts, *Contributions to Mineralogy and Petrology*, *149*(1), 113-128, doi:10.1007/s00410-004-0641-8.
Mallik, A., and R. Dasgupta (2012), Reaction between MORB-eclogite derived melts and fertile peridotite and generation of ocean island basalts, *Earth Planet. Sci. Lett.*, *329*, 97-108, doi:10.1016/j.epsl.2012.02.007.
Manga, M. (2010), Low-viscosity mantle blobs are sampled preferentially at regions of surface divergence and stirred rapidly into the mantle, *Phys. Earth Planet. Inter.*, *180*(1), 104-107.

Marcq, E. (2012), A simple 1-D radiative-convective atmospheric model designed for integration into coupled models of magma ocean planets, *Journal of Geophysical Research: Planets*, *117*(E1).
Maurice, M., N. Tosi, H. Samuel, A.-C. Plesa, C. Hüttig, and D. Breuer (2017), Onset of solid-state mantle convection and mixing during magma ocean solidification, *Journal of Geophysical Research: Planets*, *122*(3), 577-598, doi:10.1002/2016JE005250.
McNamara, A. K., and S. Zhong (2004), Thermochemical structures within a spherical mantle: Superplumes or piles?, *J. Geophys. Res.*, *109*(B07402), doi:10.1029/2003JB00287.
Mezger, K., V. Debaille, and T. Kleine (2013), Core Formation and Mantle Differentiation on Mars, *Space Science Reviews*, *174*(1), 27-48, doi:10.1007/s11214-012-9935-8.
Mojzsis, S. J., T. M. Harrison, and R. T. Pidgeon (2001), Oxygen-isotope evidence from ancient zircons for liquid water at the Earth's surface 4,300 Myr ago, *Nature*, *409*(6817), 178-181.
Moresi, L., S. Zhong, and M. Gurnis (1996), The accuracy of finite element solutions of Stokes' flow with strongly varying viscosity, *Phys. Earth Planet. Int.*, *97*(1-4), 83-94.
Mosca, I., L. Cobden, A. Deuss, J. Ritsema, and J. Trampert (2012), Seismic and mineralogical structures of the lower mantle from probabilistic tomography, *J. Geophys. Res.*, *117*, B06304, doi:10.1029/2011jb008851.
Mosenfelder, J. L., P. D. Asimow, D. J. Frost, D. C. Rubie, and T. J. Ahrens (2009), The MgSiO3 system at high pressure: Thermodynamic properties of perovskite, postperovskite, and melt from global inversion of shock and static compression data, *Journal of Geophysical Research: Solid Earth*, *114*(B1).
Mostefaoui, S., G. Lugmair, and P. Hoppe (2005), 60Fe: a heat source for planetary differentiation from a nearby supernova explosion, *The Astrophysical Journal*, *625*(1), 271.
Moulik, P., and G. Ekström (2016), The relationships between large-scale variations in shear velocity, density, and compressional velocity in the Earth's mantle, *Journal of Geophysical Research: Solid Earth*, *121*(4), 2737-2771.
Mukhopadhyay, S. (2012), Early differentiation and volatile accretion recorded in deep-mantle neon and xenon, *Nature*, *486*(7401), 101-U124, doi:10.1038/nature11141.
Mundl, A., M. Touboul, M. G. Jackson, J. M. D. Day, M. D. Kurz, V. Lekic, R. T. Helz, and R. J. Walker (2017), Tungsten-182 heterogeneity in modern ocean island basalts, *Science*, *356*(6333), 66.
Nakagawa, T., and B. A. Buffett (2005), Mass transport mechanism between the upper and lower mantle in numerical simulations of thermochemical mantle convection with multicomponent phase changes, *Earth Planet. Sci. Lett.*, *230*, 11-27.
Nakajima, M., and D. J. Stevenson (2015), Melting and mixing states of the Earth's mantle after the Moon-forming impact, *Earth Planet. Sci. Lett.*, *427*, 286-295, doi:10.1016/j.epsl.2015.06.023.
Nomura, R., H. Ozawa, S. Tateno, K. Hirose, J. Hernlund, S. Muto, H. Ishii, and N. Hiraoka (2011), Spin crossover and iron-rich silicate melt in the Earth/'s deep mantle, *Nature*, *473*(7346), 199-202.
Persh, S. E., and J. E. Vidale (2004), Reflection properties of the core-mantle boundary from global stacks of PcP and ScP, *Journal of Geophysical Research B: Solid Earth*, *109*(4), B04309 04301-04311, doi:10.1029/2003JB002768.
Plesa, A. C., N. Tosi, and D. Breuer (2014), Can a fractionally crystallized magma ocean explain the thermo-chemical evolution of Mars?, *Earth Planet. Sci. Lett.*, *403*, 225-235, doi:10.1016/j.epsl.2014.06.034.
Rizo, H., R. J. Walker, R. W. Carlson, M. Touboul, M. F. Horan, I. S. Puchtel, M. Boyet, and M. T. Rosing (2016), Early Earth differentiation investigated through 142Nd, 182W, and highly siderophile element abundances in samples from Isua, Greenland, *Geochimica et Cosmochimica Acta*, *175*, 319-336, doi:10.1016/j.gca.2015.12.007.
Rubie, D. C., S. A. Jacobson, A. Morbidelli, D. P. O'Brien, E. D. Young, J. de Vries, F. Nimmo, H. Palme, and D. J. Frost (2015), Accretion and differentiation of the terrestrial planets with implications for the compositions of early-formed Solar System bodies and accretion of water, *Icarus (USA)*, *248*, 89-108, doi:10.1016/j.icarus.2014.10.015.
Rufu, R., O. Aharonson, and H. B. Perets (2017), A multiple-impact origin for the Moon, *Nature Geoscience*.
Sasaki, S., and K. Nakazawa (1986), Metal-silicate fractionation in the growing Earth: Energy source for the terrestrial magma ocean, *J. Geophys. Res.*, *91*, 9231-9238.
Solomatov, V. (2015), 9.04 - Magma Oceans and Primordial Mantle Differentiation A2 - Schubert, Gerald, in *Treatise on Geophysics (Second Edition)*, edited, pp. 81-104, Elsevier, Oxford, doi:10.1016/B978-0-444-53802-4.00155-X.
Solomatov, V. S., P. Olson, and D. J. Stevenson (1993), Entrainment From a Bed Of Particles By Thermal-Convection, *Earth Planet. Sci. Lett.*, *120*(3-4), 387-393.
Solomatov, V. S., and D. J. Stevenson (1993a), Nonfractional Crystallization Of a Terrestrial Magma Ocean, *Journal Of Geophysical Research-Planets*, *98*(E3), 5391-5406.
Solomatov, V. S., and D. J. Stevenson (1993b), Suspension In Convective Layers and Style Of Differentiation Of a Terrestrial Magma Ocean, *Journal Of Geophysical Research-Planets*, *98*(E3), 5375-5390.
Solomatov, V. S. (2000), Fluid dynamics of a terrestrial magma ocean, in *Origin of the Earth and Moon*, edited by R. M. Canup and K. Righter, The University of Arizona Press, Tucson, AZ.
Stixrude, L., N. de Koker, N. Sun, M. Mookherjee, and B. B. Karki (2009), Thermodynamics of silicate liquids in the deep Earth, *Earth Planet. Sci. Lett.*, *278*(3), 226-232.
Tackley, P. J., and S. Xie (2002), The thermo-chemical structure and evolution of Earth's mantle: constraints and numerical models, *Phil. Trans. R. Soc. Lond. A*, *360*, 2593-2609.
Tackley, P. J. (2012), Dynamics and evolution of the deep mantle resulting from thermal, chemical, phase and melting effects, *Earth-Science Reviews*, *110*(1–4), 1-25, doi:10.1016/j.earscirev.2011.10.001.
Takahashi, E., and I. Kushiro (1983), Melting of a dry peridotite at high pressures and basalt magma genesis, *American Mineralogist*, *68*(9-10), 859-879.
Tateno, S., K. Hirose, and Y. Ohishi (2014), Melting experiments on peridotite to lowermost mantle conditions, *Journal of Geophysical Research: Solid Earth*, *119*(6), 4684-4694.
Tesoniero, A., L. Auer, L. Boschi, and F. Cammarano (2015), SPani, a whole-mantle VP and VS model: Implications

on thermo-chemical structure, paper presented at EGU General Assembly Conference Abstracts.
Thomas, C. W., Q. Liu, C. B. Agee, P. D. Asimow, and R. A. Lange (2012), Multi-technique equation of state for Fe2SiO4melt and the density of Fe-bearing silicate melts from 0 to 161 GPa, *Journal of Geophysical Research: Solid Earth*, *117*(B10), n/a-n/a, doi:10.1029/2012JB009403.
Tonks, W. B., and H. J. Melosh (1990), The physics of crystal settling and and suspension in a turbulent magma ocean, in *Origin of the Earth*, edited by H. E. Newsom and J. H. Jones, pp. 151-174, Oxford University Press, New York.
Tonks, W. B., and H. J. Melosh (1993), Magma ocean formation due to giant impacts, *Journal of Geophysical Research: Planets*, *98*(E3), 5319-5333, doi:10.1029/92JE02726.
Torsvik, T. H., M. A. Smethurst, K. Burke, and B. Steinberger (2006), Large igneous provinces generated from the margins of the large low-velocity provinces in the deep mantle, *Geophys. J. Int.*, doi:10.1111/j.1365-1246X.2006.03158.x.
Tosi, N., A. C. Plesa, and D. Breuer (2013), Overturn and evolution of a crystallized magma ocean: a numerical parameter study for Mars, *Journal of Geophysical Research: Planets*, *118*(7), 1512-1528.
Touboul, M., T. Kleine, B. Bourdon, H. Palme, and R. Wieler (2009), Tungsten isotopes in ferroan anorthosites: Implications for the age of the Moon and lifetime of its magma ocean, *Icarus (USA)*, *199*(2), 245-249.
Touboul, M., I. S. Puchtel, and R. J. Walker (2012), 182W Evidence for Long-Term Preservation of Early Mantle Differentiation Products, *Science*, *335*(6072), 1065.
Trampert, J., F. Deschamps, J. Resovsky, and D. Yuen (2004), Probabilistic tomography maps chemical heterogeneities throughout the lower mantle, *Science*, *306*(5697), 853-856.
Tucker, J. M., and S. Mukhopadhyay (2014), Evidence for multiple magma ocean outgassing and atmospheric loss episodes from mantle noble gases, *Earth Planet. Sci. Lett.*, *393*, 254-265, doi:10.1016/j.epsl.2014.02.050.
Turcotte, D. L., and G. Schubert (1982), *Geodynamics: Applications of Continuum Physics to Geological Problems*, Wiley, New York.
Urey, H. C. (1955), The cosmic abundances of potassium, uranium, and thorium and the heat balances of the Earth, the Moon, and Mars, *Proceedings of the National Academy of Sciences*, *41*(3), 127-144.
van der Hilst, R. D., S. Widiyantoro, and E. R. Engdahl (1997), Evidence for deep mantle circulation from global tomography, *Nature*, *386*(6625), 578-584.
van der Meer, D. G., W. Spakman, D. J. J. van Hinsbergen, M. L. Amaru, and T. H. Torsvik (2010), Towards absolute plate motions constrained by lower-mantle slab remnants, *Nature Geoscience*, *3*(1), 36-40, doi:10.1038/ngeo708.
Weis, D., M. O. Garcia, J. M. Rhodes, M. Jellinek, and J. S. Scoates (2011), Role of the deep mantle in generating the compositional asymmetry of the Hawaiian mantle plume, *Nature Geosci.*, *4*(12), 831-838, doi:10.1038/ngeo1328.
Wilde, S. A., J. W. Valley, W. H. Peck, and C. M. Graham (2001), Evidence from detrital zircons for the existence of continental crust and oceans on the Earth 4.4 Gyr ago, *Nature*, *409*(6817), 175-178.
Williams, C. D., M. Li, A. K. McNamara, E. J. Garnero, and M. C. van Soest (2015), Episodic entrainment of deep primordial mantle material into ocean island basalts, *Nat Commun*, *6*, doi:10.1038/ncomms9937.
Wood, B. J., M. J. Walter, and J. Wade (2006), Accretion of the Earth and segregation of its core, *Nature*, *441*(7095), 825-833.
Yaxley, G. M., and D. H. Green (1998), Reactions between eclogite and peridotite: Mantle refertilisation by subduction of oceanic crust, *Schweizerische Mineralogische und Petrographische Mitteilungen*, *78*(2), 243-255.
Zahnle, K., N. Arndt, C. Cockell, A. Halliday, E. Nisbet, F. Selsis, and N. H. Sleep (2007), Emergence of a habitable planet, in *Geology and Habitability of Terrestrial planets*, edited, pp. 35-78, Springer.
Zahnle, K., L. Schaefer, and B. Fegley (2010), Earth's earliest atmospheres, *Cold Spring Harbor perspectives in biology*, *2*(10), a004895.
Zhang, J., and C. Herzberg (1994), Melting experiments on anhydrous peridotite KLB-1 from 5.0 to 22.5 GPa, *Journal of Geophysical Research: Solid Earth*, *99*(B9), 17729-17742.
Zhong, S., M. T. Zuber, L. Moresi, and M. Gurnis (2000), Role of temperature-dependent viscosity and surface plates in spherical shell models of mantle convection, *J. Geophys. Res.*, *105*(B5), 11063-11082.